\documentclass[osajnl,onecolumn,showpacs,superscriptaddress,10pt]{revtex4-1}
\usepackage{amsmath,amssymb,graphicx,subfigure}

\begin{document}

\title{Gap solitons attached to a gapless layer}
\author{Thawatchai Mayteevarunyoo}
\email{Corresponding author: thawatch@mut.ac.th}
\affiliation{Department of Telecommunication Engineering, Mahanakorn
University of Technology, Bangkok 10530, Thailand}
\author{Boris A. Malomed}
\affiliation{Department of Physical Electronics, School of Electrical
Engineering, Faculty of Engineering, Tel Aviv University, Tel Aviv 69978,
Israel}


\begin{abstract}
We consider linear and nonlinear modes pinned to a grating-free (gapless)
layer placed between two symmetric or asymmetric semi-infinite Bragg
gratings (BGs), with a possible phase shift between them, in a medium with
the uniform Kerr nonlinearity. The asymmetry is defined by a difference
between bandgap widths in the two BGs. In the linear system, exact defect
modes (DMs) are found. Composite gap solitons pinned to the central layer
are found too, in analytical and numerical forms, in the nonlinear model. In
the asymmetric system, existence boundaries for the DMs and gap solitons,
due to the competition between attraction to the gapless layer and repulsion
from the reflectivity step, are obtained analytically. Stability boundaries
for solitons in the asymmetric system are identified by means of direct
simulations. Collisions of moving BG solitons with the gapless layer are
studied too.
\end{abstract}

\maketitle



\section{Introduction and the model}

Many optical media are based on Bragg gratings (BGs), which are built as
modulations of the local refractive index with the spatial period close to
the half-wavelength of the electromagnetic field. Fiber BGs, which operate
in the temporal domain, are used in numerous applications, and are also a
subject of fundamental studies \cite{review,Kashyap}. The interplay of the
BG with the nonlinearity of the fiber material gives rise to gap solitons
\cite{SterkeSipe}-\cite{KA}, which were originally predicted as analytical
solutions \cite{Russian}-\cite{gapsol2}, and then created in experiments
\cite{experiment1}-\cite{slow}. Spatial BG solitons were predicted too,
using periodic gratings written on a planar waveguide \cite{spatial1}-\cite%
{spatial3} or in a photonic crystal \cite{spatial4}. Solitons can also be
supported by a combination of the BG with the second-harmonic-generating
\cite{PPLM}-\cite{Kovalev} or resonant (two-level) \cite{RABR,Shabtay}
nonlinearity.

The variety of modes supported by the BGs and their potential applications
can be expanded by using various superstructures created on top of regular
gratings. The technology of the creation of such complex gratings in optical
fibers became available long ago \cite{Russell}-\cite{integrated}. In
particular, supergratings give rise to additional spectral bandgaps, in
addition to the central one induced by the underlying uniform BG \cite%
{theory}. Theoretical models of the supergratings, incorporating the
nonlinearity of the material, were developed in detail \cite{theory}-\cite%
{deep2}.

Another example of the superstructure is known in the form of the Moir\'{e}
pattern, with a sinusoidal modulation imposed on the periodic variation of
the refractive index building the ordinary BG. It gives rise to a narrow
transmission band in the middle of the central gap (resembling the effect of
the electromagnetically-induced transparency), which was proposed \cite%
{Khurgin,Moire2} and implemented experimentally \cite{SlowMoire} for the
retardation of light in the BG.

It is also possible to design BG superstructures patterns in a
\textquotedblleft semi-discrete" form, i.e., as a uniform nonlinear
waveguide with periodically inserted narrow Bragg reflectors \cite{Kobi},
and as a more general scheme, which makes use of a periodic modulation of
the BG chirp and local Bragg reflectivity \cite{we}. On the other hand,
complex grating patterns can be readily written in photonic crystals and
photonic-crystal fibers \cite{PhotCryst,PhotCryst2,PhotCryst3}.

Another relevant setting is a uniform nonlinear BG with one or several local
defects (rather than defect lattices) \cite{Goodman1}-\cite{SunYatSen}. In
particular, solitons pinned to a defect represented by a $\delta $%
-functional perturbation of the local refractive index or Bragg reflectivity
can be found in an exact analytical form \cite{JOSA-B}. The delta-function
approximates a BG with a narrow embedded layer carrying large reflectivity
of the opposite sign. Interesting results are also produced by models which
include symmetric pairs of defects: repulsive ones form cavities trapping
solitons which perform shuttle oscillations, while attractive pairs give
rise to pinned states featuring spontaneous symmetry breaking \cite{Irina}.
The interaction of moving BG solitons with weak local defects can be
described by means of the perturbation theory which was developed in detail
in Refs. \cite{Tsoy2}-\cite{MakJMO}.

The nonlinear light propagation in the ``apodized" BG, i.e., one with a
spatially inhomogeneous Bragg reflectivity, obeys the system of coupled-mode
equations for amplitudes $u$ and $v$ of the right- and left-traveling waves
\cite{deep1,Tsoy1,Tsoy2,MakJMO}:

\begin{eqnarray}
i\frac{\partial u}{\partial t}+i\frac{\partial u}{\partial x}+\kappa
(x)v+\left( \left\vert v\right\vert ^{2}+\frac{1}{2}\left\vert u\right\vert
^{2}\right) u &=&0,  \notag \\
&&  \label{CME} \\
i\frac{\partial v}{\partial t}-i\frac{\partial v}{\partial x}+\kappa ^{\ast
}(x)u+\left( \left\vert u\right\vert ^{2}+\frac{1}{2}\left\vert v\right\vert
^{2}\right) v &=&0,  \notag
\end{eqnarray}%
where $t$ and $x$ are the time and coordinate normalized so that the group
velocities of the waves are $\pm 1$, ratio $2:1$ of the cross- and
self-phase-modulation coefficients in the cubic terms is a standard feature
of the Kerr effect \cite{KA}, and $\kappa (x)$ is the inhomogeneous Bragg
reflectivity, which may be complex, taking into regard a phase shift between
the incident and reflected waves (the asterisk stands for the complex
conjugate). Replacing $t$ by propagation distance $z$, the same coupled-mode
system may be realized in the spatial domain, i.e., for planar waveguides
instead of fibers. 

A local finite-width defect corresponding to
\begin{equation}
\kappa (x)=1-b~\mathrm{sech}\left( kx\right) ,  \label{Goodman}
\end{equation}%
which describes a local depression of the Bragg reflectivity, was studied in
Ref. \cite{Goodman2}. Because such a defect attracts the wave fields, it can
support localized pinned modes, including defect modes (DMs) in the
linearized version of the model. The DMs and their nonlinear extension were
constructed in Ref. \cite{Goodman2} by means of numerical methods, and the
stability of the nonlinear pinned modes was investigated. It was found that
the instability is caused by resonances of perturbations with the continuous
spectrum.

The aim of the present work is to introduce a similar but simpler and, in a
certain sense, more general, configuration, with a gapless layer placed
between two semi-infinite gratings. We define two versions of this system.
In one, the semi-infinite BGs are mutually symmetric, with a possible phase
shift ($\alpha $) between them:
\begin{equation}
\kappa (x)=\left\{
\begin{array}{c}
1,~~\mathrm{at}~~x>0, \\
0,~~\mathrm{at}~~-L<x<0, \\
\exp \left( i\alpha \right) ,~~\mathrm{at}~~x<-L,%
\end{array}%
\right.  \label{kappa}
\end{equation}%
where $|\kappa |=1$ for $\kappa \neq 0$ may be fixed by obvious rescaling,
provided that reflectivity $|\kappa |$ takes equal absolute values in both
semi-infinite gratings. The second system is composed of asymmetric BGs,
with unequal reflectivities, $\kappa =1$ and $\kappa =\cos \alpha $ (here we
set $0\leq \alpha <\pi /2$):
\begin{equation}
\kappa (x)=\left\{
\begin{array}{c}
1,~~\mathrm{at}~~x>0, \\
0,~~\mathrm{at}~~-L<x<0, \\
\cos \alpha ,~~\mathrm{at}~~x<-L.%
\end{array}%
\right.  \label{kappa2}
\end{equation}

The presence of parameter $\alpha $ in either version, based on Eq. (\ref%
{kappa}) or (\ref{kappa2}), is an essential generalization in comparison
with the model based on Eq. (\ref{Goodman}). A still more general
configuration, with a phase shift between asymmetric BGs, is possible too, \
but such system with the double complexity is beyond the scope of the
present paper. Another advantage of the defects with the simple shapes
defined by Eqs. (\ref{kappa}) and (\ref{kappa2}) is that it is much easier
to realize them in the experiment. Further, we demonstrate below that both
linear DMs and gap solitons pinned by the gapless layer can be found in an
analytical form, which is not possible in the case of the more sophisticated
profile (\ref{Goodman}). In particular, we conclude that the nonlinear
extension of the DMs fully merges into families of the pinned solitons.

It is necessary to mention that Eq. (\ref{CME}) neglects the ordinary
dispersion or diffraction in the gapless layer, in the temporal and spatial
implementations of the BG, respectively. This assumption is relevant if the
respective dispersion/diffraction length, corresponding to $L$, is much
larger than the characteristic soliton-formation length, $z_{\mathrm{sol}}$.
Note that, in physical units, $L=1$ (this scale plays a dominant role below)
corresponds to length $\sim 1$ mm, in the spatial and temporal domains
alike, and, in either case, $z_{\mathrm{sol}}$ takes values between 1 mm and
1 cm \cite{review}-\cite{SterkeSipe}, \cite{spatial1,spatial2}. With the
carrier wavelength $\simeq 1.5$ $\mathrm{\mu }$m, $L\sim 1$ mm corresponds
to diffraction length $\sim 10$ cm $\gg z_{\mathrm{sol}}$. In the temporal
domain, the disparity between the characteristic dispersion length and $z_{%
\mathrm{sol}}$ is sill larger.

The diffraction in the gapless region should be taken into regard in the
limit of $L\rightarrow \infty $, which actually corresponds to an interface
(surface) separating a semi-infinite BG and the uniform medium. The
interaction of \textit{gap surface solitons} with such an interface was
studied before \cite{Barcelona1}-\cite{Barcelona3}.

The subsequent presentation is structured as follows. Analytical results are
presented in Section II. They include exact solutions for linear DMs in both
models (\ref{kappa}) and (\ref{kappa2}), a perturbative treatment of its
weakly nonlinear extension (explicitly presented for the symmetric system,
with $\alpha =0$), which is well corroborated by numerical results, and an
implicit analytical solution for solitons pinned to the gapless layer.
Section III reports basic numerical results for the nonlinear asymmetric
system, based on Eq. (\ref{kappa2}), including its symmetric limit, with $%
\alpha =0$. The numerical results display existence and stability areas for
the pinned solitons in the underlying parameter space, and simulations of
collisions of moving solitons with the gapless layer (for another type of
the defect, collisions were simulated in Refs. \cite{Goodman1} and \cite%
{JOSA-B}). In particular, the collision may generate a trapped soliton from
a part of the energy of the incident one, which can be used for the creation
of standing optical solitons (thus far, the best experimental result for
slow-light BG solitons was the creation of one moving at velocity $\approx
0.16$ speed of light in vacuum \cite{slow}).

\section{Analytical results}

\subsection{Stationary equations}

Stationary solutions of Eq. (\ref{CME}) with frequency $\omega $ are looked
for as%
\begin{equation}
\left\{ u\left( x,t\right) ,v\left( x,t\right) \right\} =e^{-i\omega
t}\left\{ U(x),V(x)\right\} ,  \label{omega}
\end{equation}%
where complex functions $U$ and $V$ satisfy the system of ordinary
differential equations:

\begin{eqnarray}
&&+i\frac{dU}{dx}+\omega U+\kappa (x)V+\left[ \left( \left\vert V\right\vert
^{2}+\frac{1}{2}\left\vert U\right\vert ^{2}\right) \right] U=0,  \notag \\
&&  \label{Solutions} \\
&&-i\frac{dV}{dx}+\omega V+\kappa ^{\ast }(x)U+\left[ \left( \left\vert
U\right\vert ^{2}+\frac{1}{2}\left\vert V\right\vert ^{2}\right) \right] V=0,
\notag
\end{eqnarray}%
which are compatible with the constraint reducing the two stationary fields
to one,%
\begin{equation}
U^{\ast }\left( x\right) =-V(x).  \label{UV}
\end{equation}%
The stationary solutions are characterized by the total energy, alias norm
(or total power, in terms of the spatial-domain model) and Hamiltonian,
\begin{equation}
E=\int_{-\infty }^{+\infty }\left[ \left\vert U(x)\right\vert
^{2}+\left\vert V(x)\right\vert ^{2}\right] dx,  \label{E}
\end{equation}%
\begin{gather}
H=\int_{-\infty }^{+\infty }\left\{ \left[ \frac{i}{2}\left( U\frac{dU^{\ast
}}{dx}-V\frac{dV^{\ast }}{dx}\right) +\mathrm{c.c.}\right] -\frac{1}{4}%
\left( \left\vert U\right\vert ^{4}+|V|^{4}+|U|^{2}|V|^{2}\right) \right\}
dx+H_{\mathrm{Bragg}},  \label{H} \\
H_{\mathrm{Bragg}}=-\int_{-\infty }^{+\infty }\left[ \kappa (x)U^{\ast }V+%
\mathrm{c.c.}\right] dx,  \label{int}
\end{gather}%
where $\mathrm{c.c.}$, as well as $\ast $, stand for the complex conjugate.

The well-known analytical solution for the quiescent BG solitons in the
uniform grating, with $L=0$ and constant $\kappa \equiv \left\vert \kappa
\right\vert e^{i\alpha }$, corresponds to general expressions (\ref{omega})
and (\ref{UV}), with\ \cite{Russian,gapsol1,gapsol2}%
\begin{equation}
U_{\mathrm{sol}}=\sqrt{\frac{2}{3}\left\vert \kappa \right\vert }e^{i\alpha
/2}\frac{\sin \,\theta }{\cosh \left( \left\vert \kappa \right\vert x\sin
\,\theta -i\theta /2\right) }\,,  \label{sol}
\end{equation}%
\begin{equation}
\omega _{\mathrm{sol}}=\left\vert \kappa \right\vert \cos \theta .
\label{cos}
\end{equation}%
Here, $\theta $ is an intrinsic parameter of the soliton family which takes
values $0<\theta <\pi $, with the respective soliton's frequency (\ref{cos})
belonging to the bandgap of the linearized version of Eqs. \noindent (\ref%
{CME}),%
\begin{equation}
-\left\vert \kappa \right\vert <\omega <+\left\vert \kappa \right\vert .
\label{bandgap}
\end{equation}%
The total energy (\ref{E}) of this exact solution is%
\begin{equation}
E_{\mathrm{sol}}=(8/3)\theta  \label{Esol}
\end{equation}%
(note that $E_{\mathrm{sol}}$ does not explicitly depend on $\left\vert
\kappa \right\vert $). It is known that the exact solitons are stable, in
the uniform medium, at \cite{Rich}-\cite{Trillo}, \cite{Kovalev}
\begin{equation}
\theta \leq \theta _{\max }\approx 1.01\times \pi /2,  \label{stab}
\end{equation}%
i.e., in the interval of frequencies%
\begin{equation}
\left( \omega _{\mathrm{sol}}\right) _{\mathrm{cr}}\approx -0.016\leq \omega
_{\mathrm{sol}}\leq 1,  \label{crit}
\end{equation}%
cf. the full bandgap (\ref{bandgap}).

In the asymmetric system with reflectivity profile (\ref{kappa2}), the full
bandgap is reduced, per Eq. (\ref{bandgap}), from $\omega ^{2}<1$ to $\omega
^{2}<\cos ^{2}\alpha $. In the interval between the bandgaps of the two
semi-infinite gratings, i.e., at
\begin{equation}
\cos ^{2}\alpha <\omega ^{2}<1,  \label{interstitial}
\end{equation}%
the asymmetric system gives rise to states which are localized at $x>0$ but
delocalized at $x<-L$, see below.

Exact solution (\ref{sol}) features an obvious symmetry, $\left\{ U_{\mathrm{%
sol}}(-x),V_{\mathrm{sol}}(-x)\right\} =\left\{ U_{\mathrm{sol}}^{\ast
}(x),V_{\mathrm{sol}}^{\ast }(x)\right\} $. Similarly, localized modes in
the system based on Eqs. (\ref{CME}) and (\ref{kappa}) with $\alpha =0$ are
symmetric with respect to the midpoint of the gapless layer, $x=-L/2$:
\begin{equation}
\left\{ U(-x),V(-x)\right\} =\left\{ U^{\ast }(x-L),V^{\ast }(x-L)\right\} .
\label{symm}
\end{equation}%
The presence of the phase shift ($\alpha $) in Eq. (\ref{kappa}) between the
semi-infinite gratings affects this relation (see below), but the
corresponding intensity profile remains symmetric:%
\begin{equation}
\left\vert U(-x)\right\vert ^{2}+\left\vert V(-x)\right\vert ^{2}=\left\vert
U(x-L)\right\vert ^{2}+\left\vert V(x-L)\right\vert ^{2}.  \label{density}
\end{equation}

\subsection{The linear defect mode}

\subsubsection{The symmetric system}

The linearized version of Eq. (\ref{Solutions}), subject to reduction (\ref%
{UV}), reduces to the single linear equation,%
\begin{equation}
i\frac{dU}{dx}+\omega U-\kappa (x)U^{\ast }=0.  \label{U}
\end{equation}%
This equation, with $\kappa (x)$ taken as per Eq. (\ref{kappa}) for the
symmetric system, including the phase shift, gives rise to an exact\textit{\
composite} (three-layer) DM solution with eigenfrequency $\omega _{\mathrm{%
eigen}}$:
\begin{equation}
U(x)\equiv \varepsilon U_{1}(x)=\varepsilon \left\{
\begin{array}{c}
\exp \left[ \frac{i}{2}\cos ^{-1}\left( \omega _{\mathrm{eigen}}\right) -%
\sqrt{1-\omega _{\mathrm{eigen}}^{2}}x\right] ,~\mathrm{at~~}x\geq 0, \\
B_{\mathrm{DM}}\exp \left( i\omega _{\mathrm{eigen}}\left( x+\frac{L}{2}%
\right) \right) ,~\mathrm{at~~}-L\leq x\leq 0, \\
\exp \left[ \frac{i}{2}\alpha -\frac{i}{2}\cos ^{-1}\left( \omega _{\mathrm{%
eigen}}\right) +\sqrt{1-\omega _{\mathrm{eigen}}^{2}}\left( x+L\right) %
\right] ,~\mathrm{at~~}x\leq -L,%
\end{array}%
\right.  \label{U(x)}
\end{equation}%
where $\varepsilon $ is an infinitesimal amplitude which underlies the
linearization. Relative amplitude $B_{\mathrm{DM}}$ and the eigenfrequency
[it always falls into bandgap (\ref{bandgap})] are determined \ by the
condition of the continuity of $U(x)$ at $x=0$ and $x=-L$:%
\begin{equation}
B_{\mathrm{DM}}=\exp \left( i\alpha /4+i\pi n\right) ,  \label{A}
\end{equation}%
\begin{equation}
\omega _{\mathrm{eigen}}L+\left( \alpha /2\right) =\cos ^{-1}\left( \omega _{%
\mathrm{eigen}}\right) +2\pi n,  \label{eigen}
\end{equation}%
where $n$ is an arbitrary integer. Note that the DM given by Eq. (\ref{U(x)}%
) is fully symmetric about the midpoint, as per definition (\ref{symm}), at $%
\alpha =0$, and satisfies symmetry relation (\ref{density}) at any $\alpha $%
. The respective integral energy (\ref{E}) is $E_{\mathrm{DM}}=\varepsilon
^{2}\left( L+1/\sqrt{1-\omega _{\mathrm{eigen}}^{2}}\right) $.

For a weak gapless layer, i.e., $L\rightarrow 0$ and $\alpha \rightarrow 0$,
Eq. (\ref{eigen}) for the fundamental DM (with $n=0$) yields
\begin{equation}
\omega _{\mathrm{eigen}}\approx 1-\left( 1/8\right) \left( \alpha +2L\right)
^{2},  \label{small-L}
\end{equation}%
with solution (\ref{U(x)}) degenerating into a constant at $L=\alpha =0$. On
the other hand, at $\alpha \rightarrow \pi $, Eq. (\ref{eigen}) with $n=0$
yields a vanishingly small eigenfrequency,%
\begin{equation}
\omega _{\mathrm{eigen}}\approx \frac{1}{2}\frac{\pi -\alpha }{1+L}.
\label{alpha-pi}
\end{equation}%
Further, the largest possible eigenfrequency, $\omega _{\mathrm{eigen}}=1$,
is attained by the solution to Eq. (\ref{eigen}) for the fundamental DM\ ($%
n=0$) at $\alpha \left( \omega _{\mathrm{eigen}}=1\right) =-2L$, and close
to this point the solution takes the same form as given by Eq. (\ref{small-L}%
). The fundamental DM does not exist at%
\begin{equation}
\alpha <\alpha _{\min }\equiv -2L.  \label{min}
\end{equation}

The relation between the eigenfrequency and $\alpha $, produced by Eq. (\ref%
{eigen}) with $n=0$, is displayed, for $L=0.5$ and $L=2.0$, in Fig. \ref%
{figure1} [$\omega <0$ corresponds to the DM subject to constraint $%
U(x)=+V^{\ast }(x)$, instead of Eq. (\ref{UV})], which demonstrates that the
relation keeps its qualitative shape at all relevant values of the
gapless-layer's width. In particular, in the limit of $L\rightarrow 0$,
which corresponds to the junction of two semi-infinite gratings with phase
jump $\alpha $ between them (here, we can fix $0\leq \alpha \leq \pi $),
straightforward consideration of Eq. (\ref{eigen}) yields%
\begin{equation}
\omega _{\mathrm{eigen}}\left( L\rightarrow 0\right) =\cos \left( \alpha
/2\right) -\left( L/2\right) \sin \alpha +O\left( L^{2}\right) ,  \label{L=0}
\end{equation}%
the expansion of which at $\alpha \rightarrow 0$ complies with Eq. (\ref%
{small-L}). At $\alpha =0$, characteristic eigenfrequencies are
\begin{equation}
\omega _{\mathrm{eigen}}\left( L=0.5\right) \approx 0.90,~\omega _{\mathrm{%
eigen}}\left( L=1.0\right) \approx 0.74,~\omega _{\mathrm{eigen}}\left(
L=1.5\right) \approx 0.61,\omega _{\mathrm{eigen}}\left( L=2.0\right)
\approx 0.51.\allowbreak  \label{4-L}
\end{equation}

\begin{figure}[tbp]
\centering\subfigure[]{\includegraphics[width=3in]{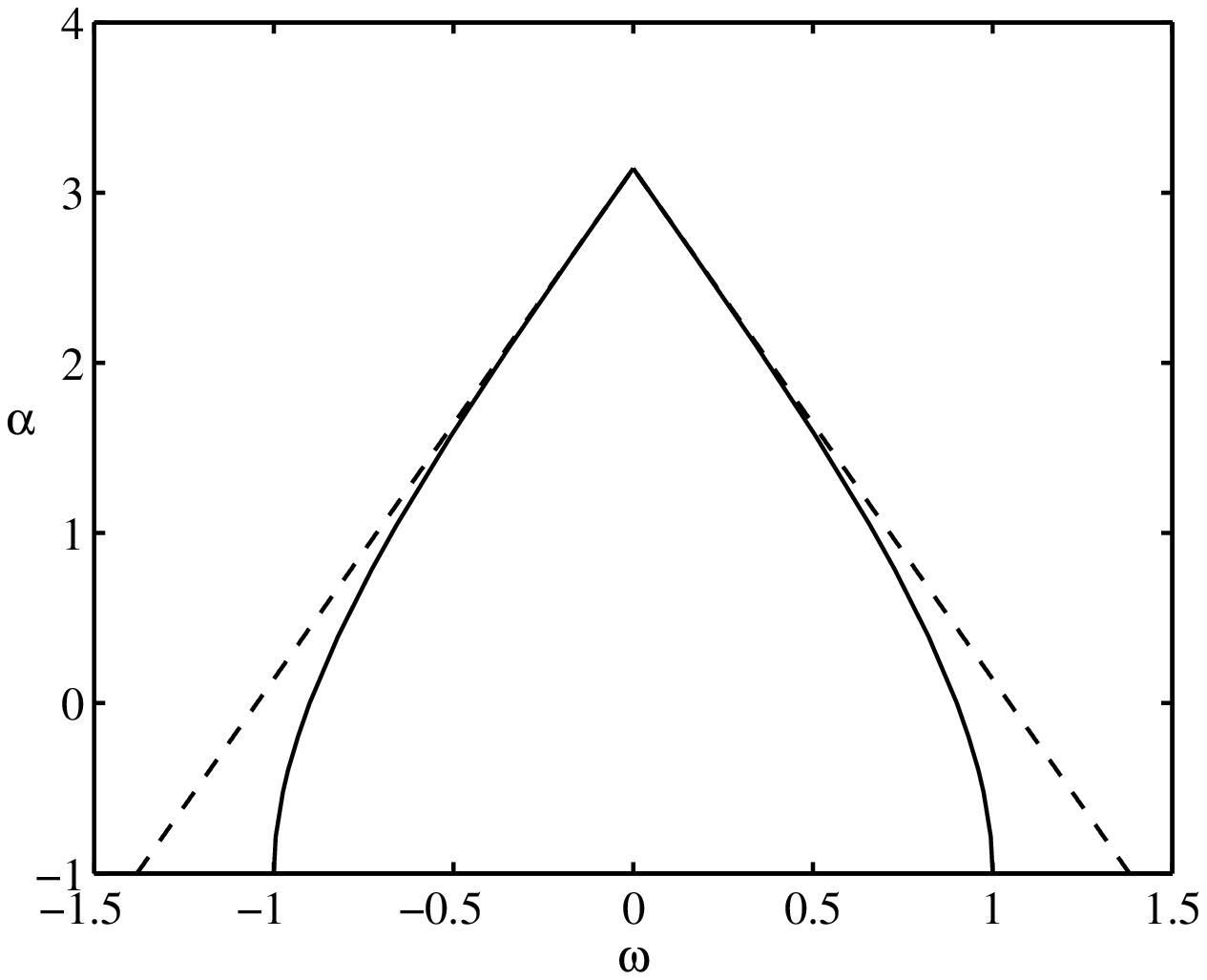}}%
\subfigure[]{\includegraphics[width=3in]{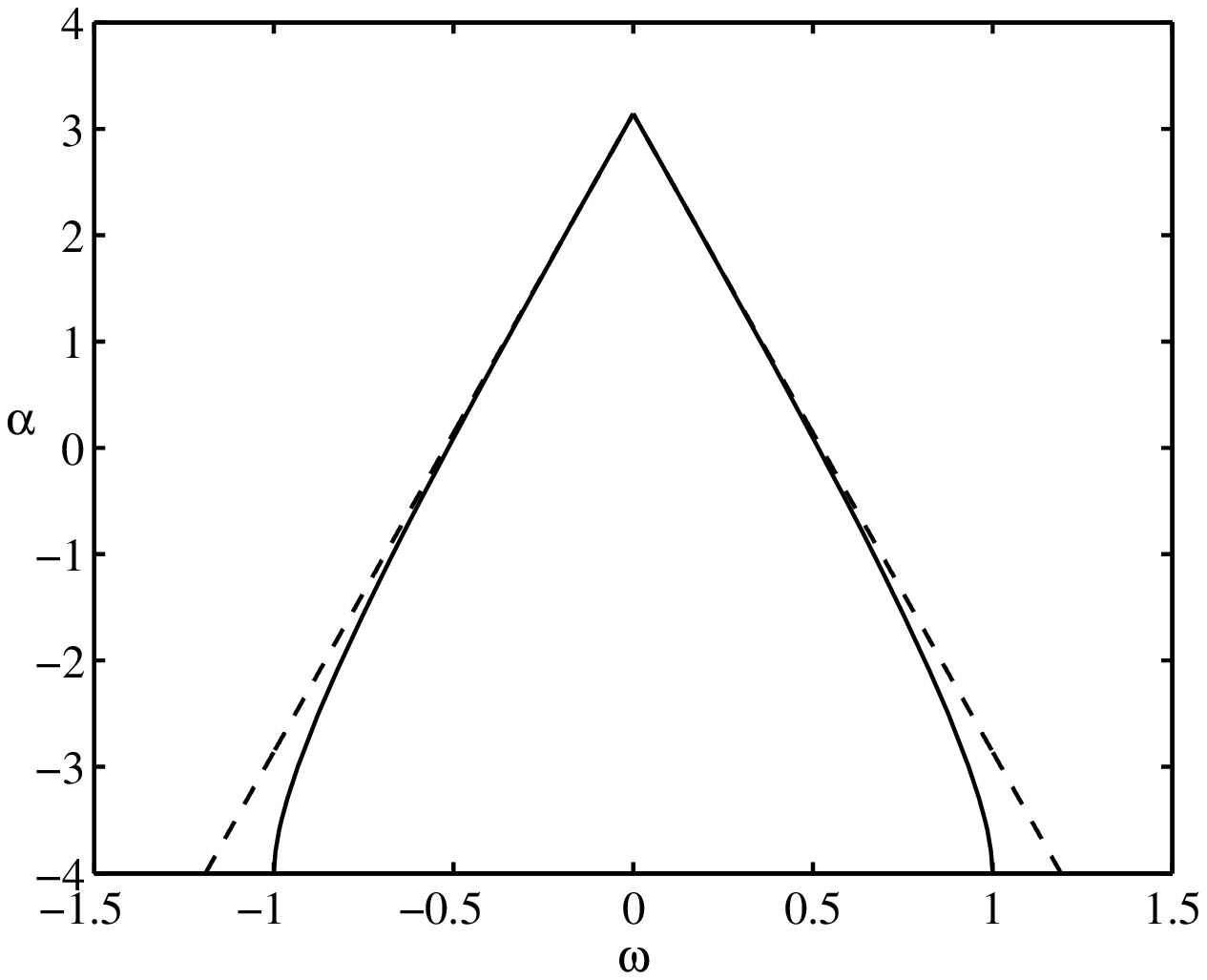}}
\caption{Relations between the eigenfrequency of the fundamental defect
mode, in the linear symmetric system with phase shift $\protect\alpha $
[defined per Eq. (\protect\ref{kappa})], and $\protect\alpha $, as produced
by Eq. (\protect\ref{eigen}) with $n=0$, at fixed values of the width of the
gapless layer: (a) $L=0.5$; (b) $L=2.0$. The dashed straight lines depict
analytical approximation (\protect\ref{alpha-pi}).}
\label{figure1}
\end{figure}

At large values of $L$, higher-order DMs appear, corresponding to $n>0$ in
Eq. (\ref{eigen}) [higher-order linear DMs were also found numerically in
the model including the defect profile (\ref{Goodman}) \cite{Goodman2}].
Indeed, higher-order DMs emerge, for each $n\geq 1$, at $\omega =1$, i.e., at%
\begin{equation}
L_{\mathrm{cr}}^{(n)}=2\pi n-\alpha /2\text{.}  \label{Ln}
\end{equation}%
For instance, at $\alpha =0$ and with $L$ slightly exceeding the first
critical value (\ref{Ln}), i.e., $L=2\pi +l$, $0<l\ll 2\pi $, the emerging
eigenfrequency is obtained as $\omega _{\mathrm{eigen}}^{(n=1)}\approx
1-l^{2}/2,$ cf. Eq. (\ref{small-L}), while its fundamental counterpart,
found as a numerical solution of Eq. (\ref{eigen}), is $\omega _{\mathrm{%
eigen}}^{(n=0)}(L=2\pi )\approx 0.22$. As follows from Eq. (\ref{Ln}), for
given $L$ the total number of defect modes is%
\begin{equation}
n_{\mathrm{DM}}=1+\left[ \frac{L+\left( \alpha /2\right) }{2\pi }\right] ,
\label{nDM}
\end{equation}%
where $[...]$ denotes the integer part.

Below, reporting numerical results for the full nonlinear system based on
Eqs. (\ref{CME}) and (\ref{kappa}), we focus on moderately wide gapless
layers, with $L\leq 2$, and $0\leq \alpha \leq \pi /3,$ for which there is
the single DM, corresponding to $n=0$.

Outside of bandgap (\ref{bandgap}), an exact composite solution of the
linearized equation (\ref{U}) for delocalized states can be found too [cf.
Eq. (\ref{U(x)})]:%
\begin{equation}
U(x)=\varepsilon \left\{
\begin{array}{c}
A_{+}\exp \left( i\sqrt{\omega ^{2}-1}x\right) +A_{-}\exp \left( -i\sqrt{%
\omega ^{2}-1}x\right) ,~\mathrm{at~~}x\geq 0, \\
\exp \left( i\omega \left( x+\frac{L}{2}\right) \right) ,~\mathrm{at~~}%
-L\leq x\leq 0, \\
C_{+}\exp \left( i\sqrt{\omega ^{2}-1}\left( x+L\right) \right) +C_{-}\exp
\left( -i\sqrt{\omega ^{2}-1}\left( x+L\right) \right) ,~\mathrm{at~~}x\leq
-L,%
\end{array}%
\right.  \label{deloc}
\end{equation}%
where the amplitudes are%
\begin{eqnarray}
A_{+} &=&\frac{e^{\frac{i}{2}\omega L}-e^{-\frac{i}{2}\omega L}\left( \omega
-\sqrt{\omega ^{2}-1}\right) }{2\sqrt{\omega ^{2}-1}\left( \omega +\sqrt{%
\omega ^{2}-1}\right) },~A_{-}=\left( \omega -\sqrt{\omega ^{2}-1}\right)
A_{+}^{\ast }~,  \notag \\
C_{+} &=&\frac{e^{-\frac{i}{2}\omega L}-e^{\frac{i}{2}\omega L+i\alpha
}\left( \omega -\sqrt{\omega ^{2}-1}\right) }{2\sqrt{\omega ^{2}-1}\left(
\omega +\sqrt{\omega ^{2}-1}\right) },~C_{-}=\left( \omega -\sqrt{\omega
^{2}-1}\right) e^{i\alpha }C_{+}^{\ast }~.  \label{CC}
\end{eqnarray}%
The existence of this solution at all $\omega ^{2}\geq 1$ corroborates that
the insertion of the symmetric gapless layer does not make bandgap (\ref%
{bandgap}) broader than in the uniform grating, including the case when the
phase shift is present, $\alpha \neq 0$.

\subsubsection{The asymmetric system}

The exact three-layer solution (\ref{U(x)})-(\ref{eigen}) for the linear DM
can be easily generalized for the system with asymmetric reflectivity
profile (\ref{kappa2}):%
\begin{equation}
U(x)\equiv \varepsilon U_{1}(x)=\varepsilon \left\{
\begin{array}{c}
\exp \left[ \frac{i}{2}\cos ^{-1}\left( \omega _{\mathrm{eigen}}\right) -%
\sqrt{1-\omega _{\mathrm{eigen}}^{2}}x\right] ,~\mathrm{at~~}x\geq 0, \\
\tilde{B}_{\mathrm{DM}}\exp \left( i\omega _{\mathrm{eigen}}\left( x+\frac{L%
}{2}\right) \right) ,~\mathrm{at~~}-L\leq x\leq 0, \\
\exp \left[ -\frac{i}{2}\cos ^{-1}\left( \frac{\omega _{\mathrm{eigen}}}{%
\cos \alpha }\right) +\sqrt{\cos ^{2}\alpha -\left( \omega _{\mathrm{eigen}%
}\right) ^{2}}\left( x+L\right) \right] ,~\mathrm{at~~}x\leq -L,%
\end{array}%
\right.  \label{U(x)2}
\end{equation}%
\begin{equation}
\tilde{B}_{\mathrm{DM}}=\exp \left[ \frac{i}{2}\left( \cos ^{-1}\left(
\omega _{\mathrm{eigen}}\right) -\omega _{\mathrm{eigen}}L\right) \right] ,
\label{A2}
\end{equation}%
and Eq. (\ref{eigen}) for the eigenfrequency is replaced by
\begin{equation}
\omega _{\mathrm{eigen}}L=\frac{1}{2}\left[ \cos ^{-1}\left( \omega _{%
\mathrm{eigen}}\right) +\cos ^{-1}\left( \frac{\omega _{\mathrm{eigen}}}{%
\cos \alpha }\right) \right] +2\pi n.  \label{eigen-asymm}
\end{equation}

In the symmetric system, the DM exists at all values of width $L$ of the
gapless layer, including $L=0$ [see Eq. (\ref{L=0})], provided that the
phase shift satisfies condition $\alpha >-2L$, see Eq. (\ref{min}). The
situation is drastically different in the asymmetric system, because, while
the gapless layer attracts the DM (or soliton, see below), the reflectivity
step, accounted for by $\cos \alpha <1$ in Eq. (\ref{kappa2}), repels the
localized mode. The result of the competition of the attraction and
repulsion is that, for given $\alpha $, a solution of Eq. (\ref{eigen-asymm}%
) exists only if $L$ exceeds a minimum value,%
\begin{equation}
L\geq L_{\min }^{\mathrm{(DM)}}=\frac{\alpha +4\pi n}{2\cos \alpha },
\label{Lmin}
\end{equation}%
while at $L<L_{\min }$ the DM does not exist. Exactly at $L=L_{\min }$, the
eigenfrequency coincides with the edge of the respective bandgap (\ref%
{bandgap}), $\omega _{\mathrm{eigen}}=\cos \alpha $. Further, it follows
from Eq. (\ref{Lmin}) that the total number of DMs in the asymmetric system
is%
\begin{equation*}
\tilde{n}_{\mathrm{DM}}=\left[ 1+\frac{2L\cos \alpha -\alpha }{4\pi }\right]
,
\end{equation*}%
cf. the similar result (\ref{nDM}) for the symmetric model. Another
corollary of Eq. (\ref{Lmin}) is that, for fixed $L$, the fundamental DM
(the one with $n=0$) in the asymmetric system exists in interval%
\begin{equation}
0\leq \alpha \leq \alpha _{\max },  \label{max}
\end{equation}%
with $\alpha _{\max }$ defined by equation%
\begin{equation}
2L\cos \left( \alpha _{\max }\right) =\alpha _{\max }~.  \label{alpha}
\end{equation}%
At small and large $L$, solutions to Eq. (\ref{alpha}) are%
\begin{equation}
\alpha _{\max }\approx \left\{
\begin{array}{c}
2L-4L^{3},~\mathrm{at}~~L\rightarrow 0, \\
\left( \pi /2\right) \left[ 1-1/\left( 2L\right) \right] ,~~\mathrm{at}%
~~L\rightarrow \infty .%
\end{array}%
\right.  \label{approx}
\end{equation}%
Particular solutions are$\allowbreak $
\begin{equation}
\alpha _{\max }\left( L=0.5\right) \approx 0.74,\alpha _{\max }\left(
L=1.0\right) \approx 1.03,\alpha _{\max }\left( L=1.5\right) \approx
1.17,\alpha _{\max }\left( L=2.0\right) \approx 1.25.\allowbreak
\label{part}
\end{equation}%
%
%
%
%
%

\begin{figure}[tbp]
\caption{(Color online) The relation between the eigenvalue of the
fundamental linear defect mode in the asymmetric system and asymmetry
parameter $\protect\alpha $, as per Eq. (\protect\ref{eigen-asymm}) with $n=0
$, at fixed values of width $L$ of the gapless layer: $L=0.5$, $L=1.0$, $%
L=1.5$, and $L=2.0$. The eigenvalues are shown in their existence region, $%
0\leq \protect\alpha \leq \protect\alpha _{\max }(L)$, see the text. The
dashed green curve shows the boundary of bandgap (\protect\ref{bandgap}), $%
\protect\omega =\cos \protect\alpha $, in the same region.}\centering%
\includegraphics[width=3in]{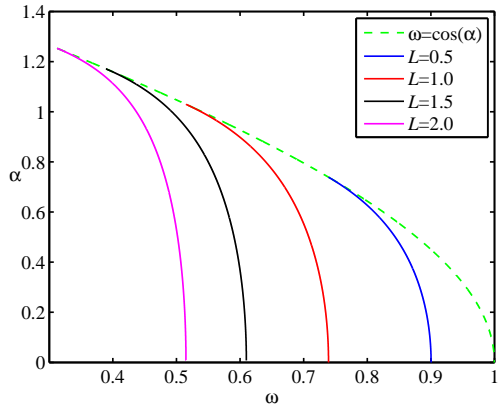}
\label{extra_fig2}
\end{figure}

Finally, for given $L$, the DM's eigenfrequency, $\omega _{\mathrm{eigen}}$,
decreases from its value in the symmetric system [the one determined by Eq. (%
\ref{eigen}) with $\alpha =0$] to the above-mentioned value, $\cos \left(
\alpha _{\max }(L)\right) $, with the increase of the asymmetry parameter, $%
\alpha $, from zero to $\alpha _{\max }(L)$. These dependences are displayed
in Fig. \ref{extra_fig2}, which can be compared to their counterparts in the
symmetric system with phase shift $\alpha $, shown in Fig. \ref{figure1}.

As mentioned above, in interval (\ref{interstitial}) between the broad and
reduced bandgaps the asymmetric system supports semi-delocalized linear
modes. The explicit form of these composite solutions is%
\begin{equation*}
U(x)=\varepsilon \left\{
\begin{array}{c}
\exp \left( \frac{i}{2}\cos ^{-1}\omega -\sqrt{1-\omega ^{2}}x\right) ,~%
\mathrm{at~~}x\geq 0, \\
\exp \left[ i\left( \frac{1}{2}\cos ^{-1}\omega +\omega x\right) \right] ,~%
\mathrm{at~~}-L\leq x\leq 0, \\
\tilde{C}_{+}\exp \left( i\sqrt{\omega ^{2}-\cos ^{2}\alpha }\left(
x+L\right) \right) +\tilde{C}_{-}\exp \left( -i\sqrt{\omega ^{2}-\cos
^{2}\alpha }\left( x+L\right) \right) ,~\mathrm{at~~}x\leq -L,%
\end{array}%
\right.
\end{equation*}%
where the amplitudes are%
\begin{eqnarray*}
\tilde{C}_{+} &=&\frac{\left( \omega +\sqrt{\omega ^{2}-\cos ^{2}\alpha }%
\right) e^{i\phi }-\left( \cos \alpha \right) e^{-i\phi }}{\sqrt{\omega
^{2}-\cos ^{2}\alpha }},~\tilde{C}_{-}=\frac{\omega -\sqrt{\omega ^{2}-\cos
^{2}\alpha }}{\cos \alpha }C_{+}^{\ast }~, \\
\phi &\equiv &\frac{1}{2}\cos ^{-1}\omega -L\omega .
\end{eqnarray*}%
cf. Eqs. (\ref{deloc}), (\ref{CC}) and (\ref{U(x)2}).

\subsection{The nonlinear correction to the eigenfrequency of the defect mode%
}

If the weak nonlinearity is taken into account, assuming that $\varepsilon $
in Eqs. (\ref{U(x)}) and (\ref{U(x)}) is small but finite, in the first
approximation it gives a small correction to the eigenfrequency,%
\begin{equation}
\omega _{\mathrm{eigen}}=\omega _{\mathrm{eigen}}^{(0)}+\varepsilon
^{2}\omega _{\mathrm{eigen}}^{(2)}~\text{.}  \label{corr}
\end{equation}%
We perform further analysis of this situation for the symmetric system, with
$\alpha =0$. Multiplying the first equation\ of system (\ref{Solutions}) by
solution (\ref{U(x)}), found in the linear approximation, and performing
integration over $dx$ yields, with regard to relation (\ref{UV}),%
\begin{equation}
\omega _{\mathrm{eigen}}^{(2)}=-\frac{3}{2}\frac{\int_{-\infty }^{+\infty
}\left\vert U_{1}(x)\right\vert ^{4}dx}{\int_{-\infty }^{+\infty }\left\vert
U_{1}(x)\right\vert ^{2}dx}=-\frac{3}{2}\frac{1+2\sqrt{1-\left( \omega _{%
\mathrm{eigen}}^{(0)}\right) ^{2}}L}{1+\sqrt{1-\left( \omega _{\mathrm{eigen}%
}^{(0)}\right) ^{2}}L}.  \label{corr2}
\end{equation}

\begin{figure}[tbp]
\centering\subfigure[]{\includegraphics[width=3in]{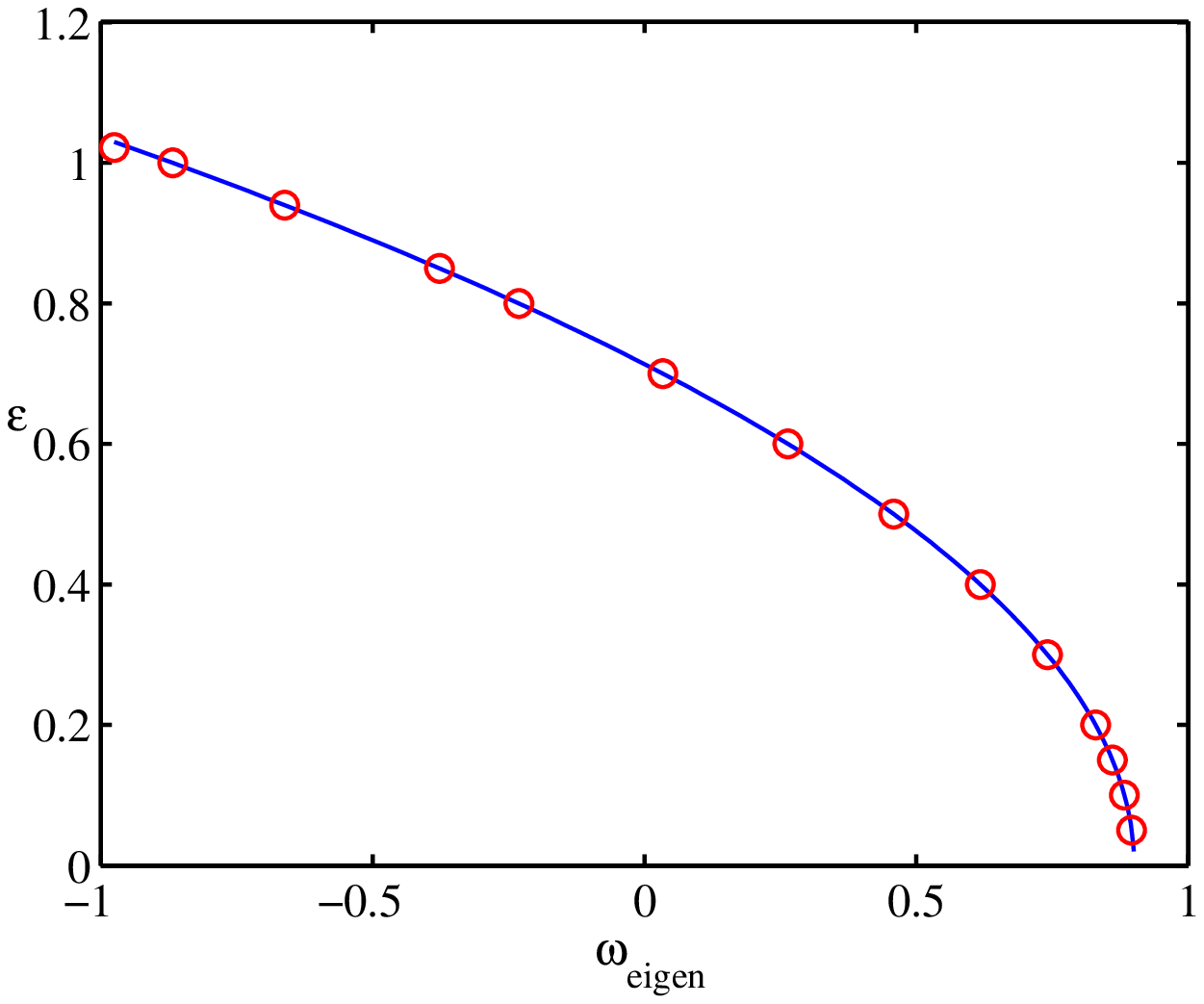}}%
\subfigure[]{\includegraphics[width=3in]{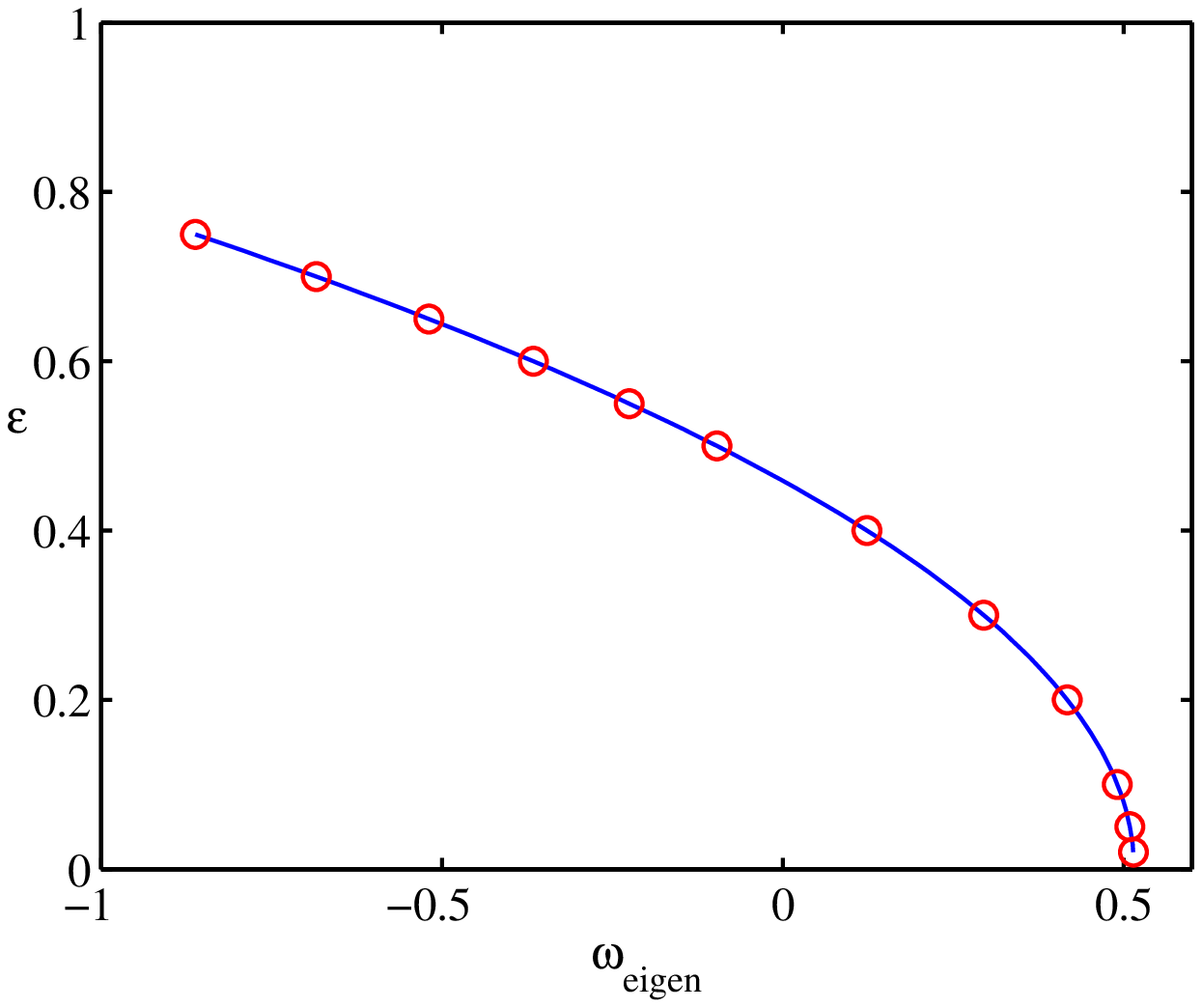}}
\caption{(Color online) Chains of circles display numerically found
eigenfrequencies of the nonlinear defect mode versus its amplitude in the
symmetric system with $\protect\alpha =0$, at different values of the width
of the gapless layer: (a) $L=0.5$; (b) $L=2.0$. Blue solid lines represent
the respective analytical approximation produced by Eqs. (\protect\ref{corr}%
) and (\protect\ref{corr2}).}
\label{figure2}
\end{figure}

Here and in the subsequent numerical analysis, we used the Newton's
iteration method to construct nonlinear DMs, using the linear modes, given
by Eqs. (\ref{U(x)}) - (\ref{eigen}), as the initial guess. The solution
domain, with periodic boundary conditions, is $|x|\leq 60$ (in all the
cases, the width of localized modes is much smaller than the size of the
domain). The resulting $\omega _{\mathrm{eigen}}(\varepsilon )$ curves for
the nonlinear mode is plotted in Fig. \ref{figure2} for two fixed values of $%
L$. The figure shows close agreement between the numerically found
eigenfrequency and the analytical prediction produced by Eqs. (\ref{corr})
and (\ref{corr2}) (the agreement is equally good at other values of $L$).

\begin{figure}[tbp]
\centering\subfigure[]{\includegraphics[width=3in]{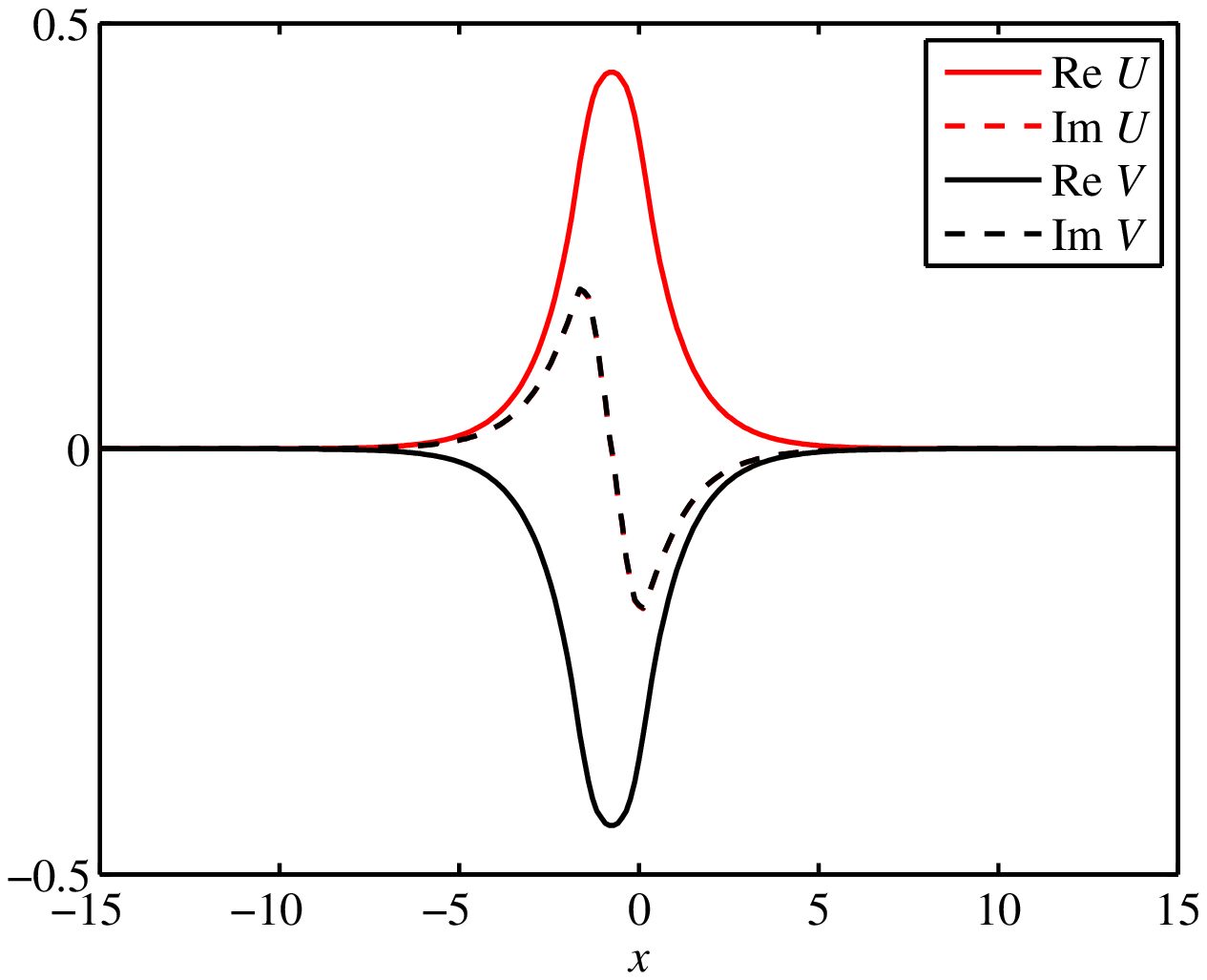}}%
\subfigure[]{\includegraphics[width=3in]{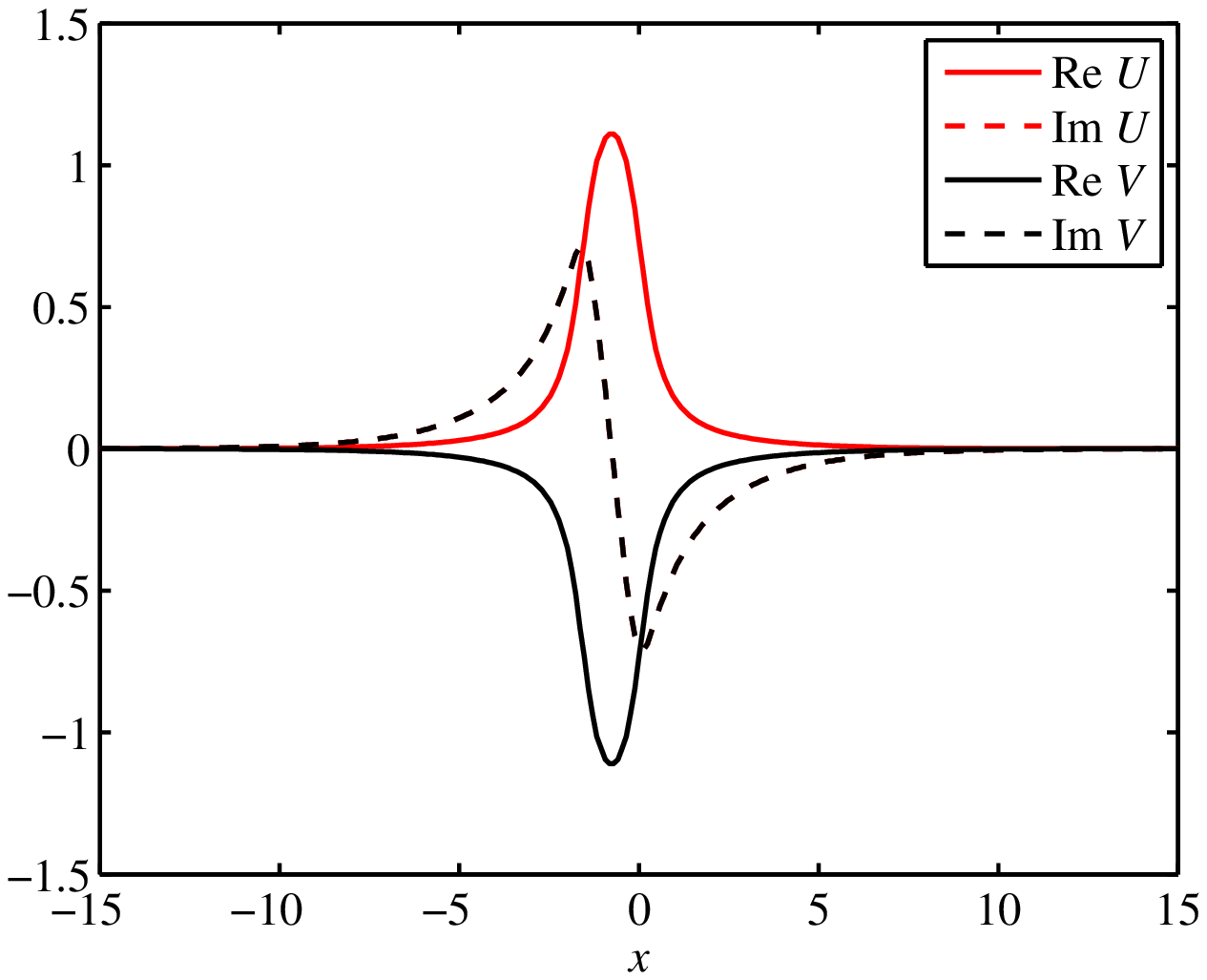}} \subfigure[]{%
\includegraphics[width=3in]{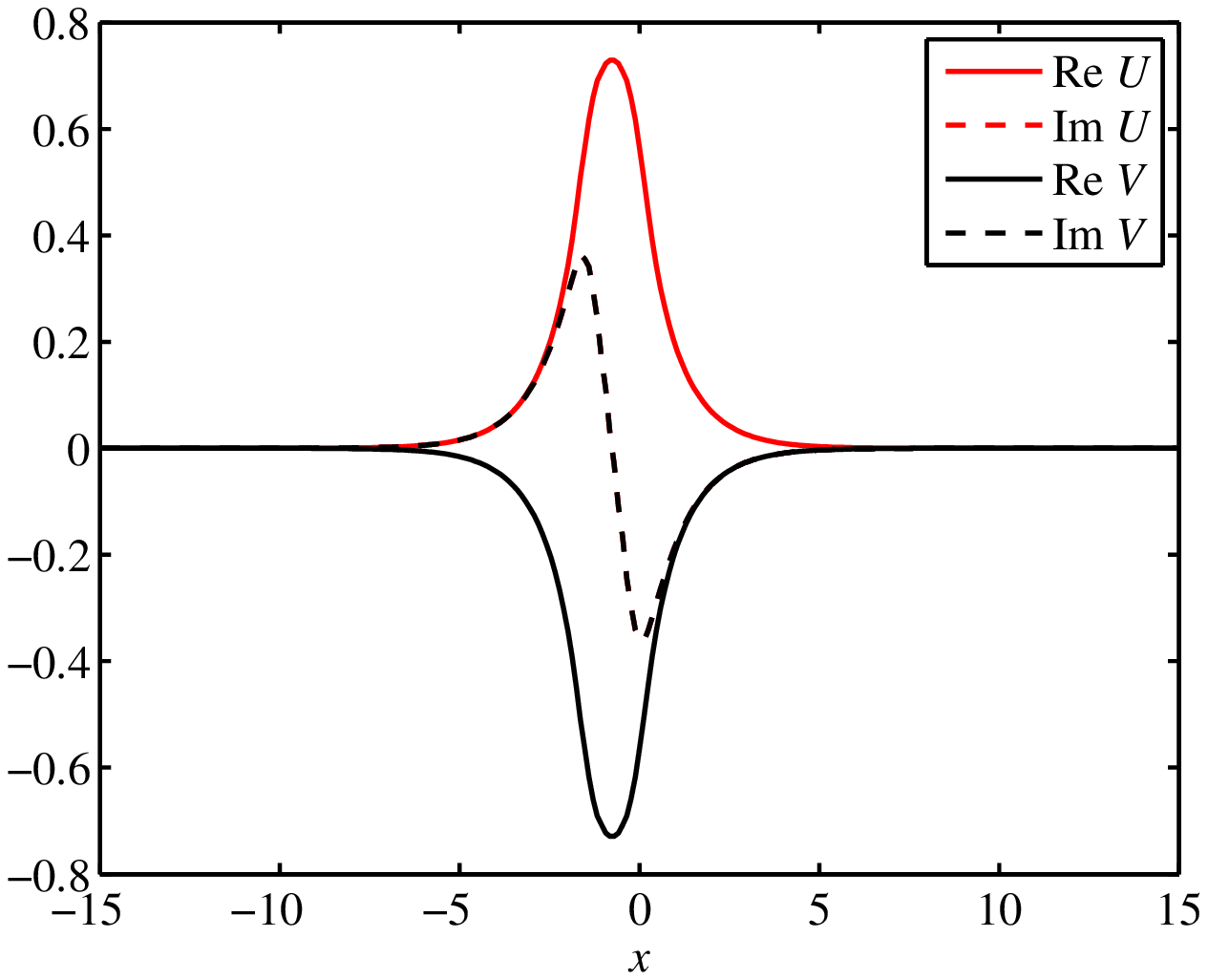}}\subfigure[]{%
\includegraphics[width=3in]{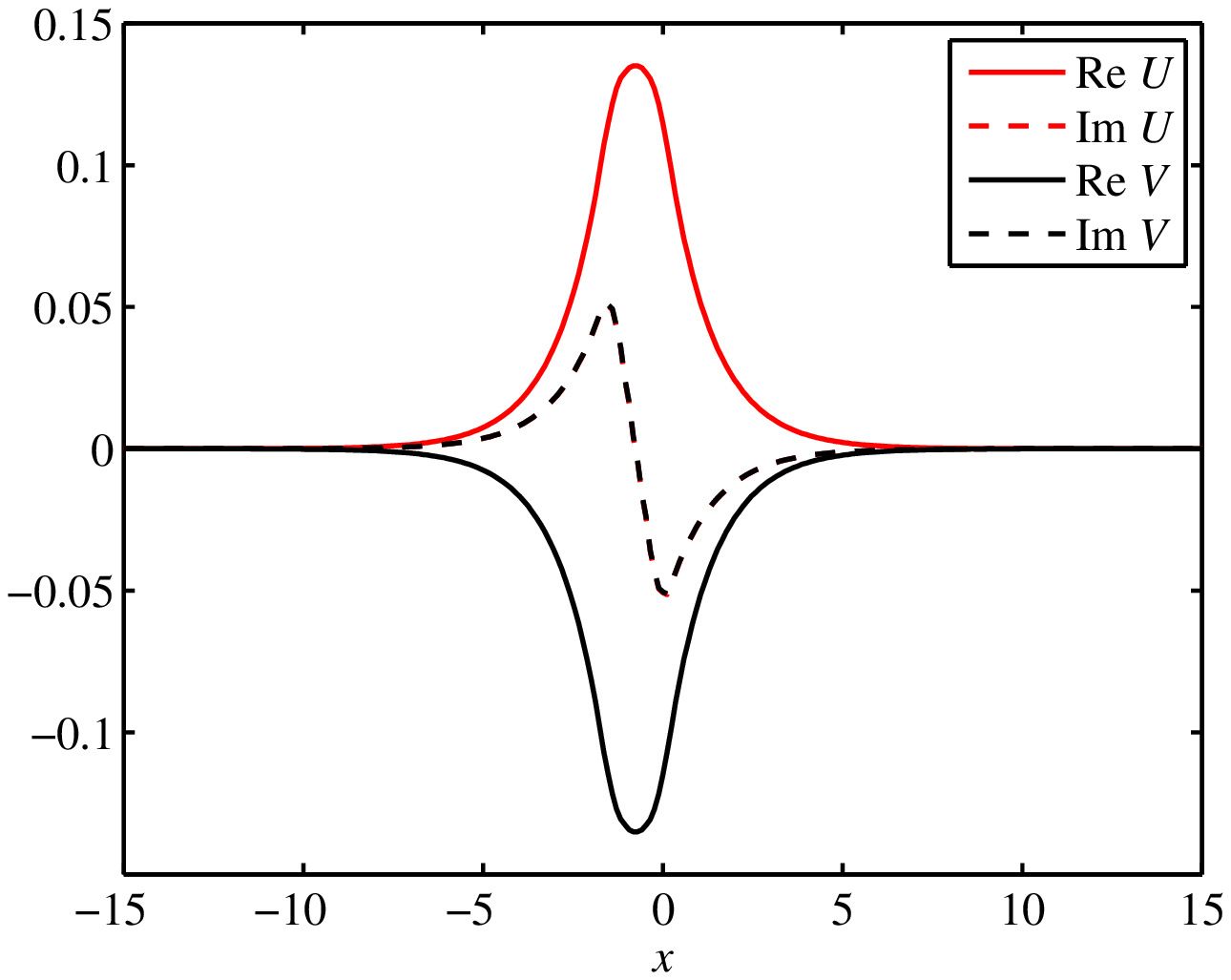}}
\caption{(Color online) Typical examples of the nonlinear and linear defect
modes in the symmetric system with $\protect\alpha =0$ and $L=1.5$: (a) $%
\protect\omega =0.4016$, $\protect\varepsilon =0.3$; (b) $\protect\omega %
=-0.8715$, $\protect\varepsilon =0.8$; (c) $\protect\omega =0$, $\protect%
\varepsilon =0.5132$; (d) $\protect\omega =0.6099$, $\protect\varepsilon =0$%
. Values of $\protect\omega $ corresponding to (a) and (b) are virtually
exactly predicted by Eq. (\protect\ref{corr2}). The linear mode in (d) is
produced by Eq. (\protect\ref{U(x)}). Because the solutions obey relation (%
\protect\ref{UV}), the curves depicting imaginary parts of $U(x)$ and $V(x)$
are identical, while their real parts have opposite signs.}
\label{figure3}
\end{figure}

Typical examples of the nonlinear DM, numerically found for $\omega >0$ and $%
\omega <0$, are shown in Figs. \ref{figure3}(a,b,c). As demonstrated below
[see Fig. \ref{figure4}(a)], the nonlinear states, i.e., as a matter of
fact, pinned gap solitons, displayed here in panels (a) and (b,c), are
stable and unstable, respectively. In addition, Fig. \ref{figure3}(d)
displays an example of the linear DM.

\subsection{Gap solitons}

\subsubsection{The symmetric system}

Stationary solutions of the nonlinear symmetric system (\ref{Solutions}),
including phase shift $\alpha $, for gap solitons pinned to the gapless
layer, which are subject to restriction (\ref{UV}), can also be constructed
as composite modes, by juxtaposing solutions (\ref{sol}) in the
semi-infinite gratings, and a plane-wave state in the gapless layer [cf.
solution (\ref{U(x)}) for the DM]:%
\begin{equation}
U(x)=\left\{
\begin{array}{c}
\sqrt{\frac{2}{3}}\frac{\sin \,\theta }{\cosh \left[ (\sin \,\theta )\left(
x+\xi \right) -i\theta /2\right] }, \\
B_{\mathrm{sol}}\exp \left[ i\left( \omega +\frac{3}{2}\left\vert B_{\mathrm{%
sol}}\right\vert ^{2}\right) \left( x+L\right) \right] , \\
\sqrt{\frac{2}{3}}e^{i\alpha /2}\frac{\sin \,\theta }{\cosh \left[ (\sin
\,\theta )\left( x+L-\xi \right) -i\theta /2\right] },%
\end{array}%
\right.  \label{soliton}
\end{equation}%
\begin{equation}
B_{\mathrm{sol}}=\sqrt{\frac{2}{3}}e^{i\alpha /2}\frac{\sin \,\theta }{\cosh
\left( \xi \sin \,\theta +i\theta /2\right) }~,  \label{B}
\end{equation}%
where $\theta $ is related to the soliton's frequency by Eq. (\ref{cos}),
and $\xi >0$ is an offset, which is determined by the condition of the
continuity of expression (\ref{soliton}) at $x=0$ [the continuity at $x=-L$
is provided by Eq. (\ref{B})]:%
\begin{equation}
\left[ \cos \theta +\frac{2\sin ^{2}\theta }{\cosh \left( 2\xi \sin \theta
\right) +\cos \theta }\right] L+\frac{\alpha }{2}=2\tan ^{-1}\left[ \tanh
\left( \xi \sin \,\theta \right) \cdot \tan \left( \frac{\theta }{2}\right) %
\right] +2\pi n,  \label{eigen2}
\end{equation}%
with arbitrary integer $n$. The linearization limit corresponds to $\xi
\rightarrow \infty $, which brings Eq. (\ref{eigen2}) back into the form of
Eq. (\ref{eigen}) for the linear DM, taking relation (\ref{cos}) into
regard. This fact implies that the nonlinear extension of the linear DM
coincides with the pinned gap solitons. In other words, it can be readily
checked that, for given frequency $\omega $ belonging to bandgap (\ref%
{bandgap}), the nonlinear DM coincides with the respective gap soliton.

It is relevant to mention that exact composite solutions were previously
obtained in other nonlinear models with multi-layer structures \cite%
{Kominis1}-\cite{Kominis3}. Most of those models are built as nonlinear
lattices of the Kronig-Penney type \cite{RMP}.

Thus, for given $\theta \equiv \cos ^{-1}\left( \omega _{\mathrm{sol}%
}\right) $, Eq. (\ref{eigen2}) determines offset $\xi $ and, through it, the
entire solution given by Eqs. (\ref{soliton}) and (\ref{B}) for the
fundamental pinned soliton at $n=0$, and higher-order ones at $n\neq 0$. In
particular, $\xi $ and $\theta $ determine the squared amplitude of the
soliton:%
\begin{equation}
\left\vert B_{\mathrm{sol}}\right\vert ^{2}=\frac{4}{3}\frac{\sin ^{2}\theta
}{\cosh \left( 2\xi \sin \theta \right) +\cos \theta }.  \label{ampl}
\end{equation}

Although the analytical solution for the pinned soliton is found here in an
implicit form, it produces some simple consequences. First, the intensity
profile corresponding to Eq. (\ref{soliton}) is obviously a \textit{flat-top}
one, as $\left\vert U(x)\right\vert ^{2}+\left\vert V(x)\right\vert
^{2}=2\left\vert B_{\mathrm{sol}}\right\vert ^{2}$ does not depend on $x$ in
the gapless layer, see Eq. (\ref{ampl}) and a typical example in Fig. \ref%
{figure6}. Second, it is obvious too that the solution obeys the symmetry
condition (\ref{density}).
\begin{figure}[tbp]
\centering
\includegraphics[width=3in]{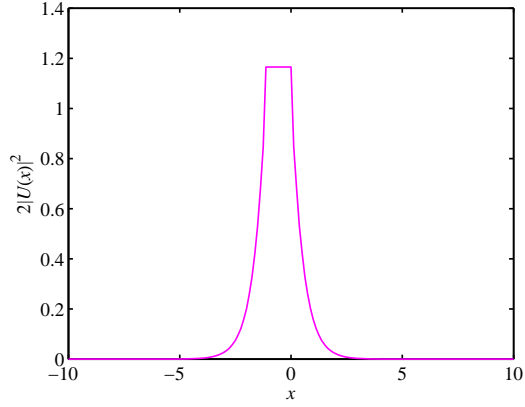}
\caption{(Color online) An example of the intensity profile of a numerically
found stable \textquotedblleft flat-top" soliton, at $\protect\alpha =0$, $%
L=1.125$, $\protect\omega =0.1$.}
\label{figure6}
\end{figure}

Note that the composite solution given by Eq. (\ref{soliton}) is \textit{%
smoothest}, i.e., without jumps of the derivative of the intensity profile, $%
\left\vert U(x)\right\vert ^{2}$, at edges of the gapless layer, $x=0$ and $%
x=-L$, if Eq. (\ref{eigen2}) yields $\xi =0$. It is easy to see that, for
the fundamental pinned soliton ($n=0$), this happens at
\begin{equation}
\omega =2+\alpha /\left( 2L\right) .  \label{smooth}
\end{equation}%
The soliton with this frequency may exist if $\omega $ falls into bandgap (%
\ref{bandgap}), i.e., at $\alpha <-2L$. Interestingly, this is precisely the
case when the linear DM does not exist, pursuant to Eq. (\ref{min}).


Lastly, in the limit of $L=0$, i.e., for the junction between two
semi-infinite gratings with phase shift $\alpha $ between them, it is easy
to find an explicit solution of Eq. (\ref{eigen2}):%
\begin{equation}
\xi =\left( 2\sin \theta \right) ^{-1}\ln \left( \frac{\tan \left( \theta
/2\right) +\tan \left( \alpha /4\right) }{\tan \left( \theta /2\right) -\tan
\left( \alpha /4\right) }\right) ,  \label{ln}
\end{equation}%
which exists for $\theta >\alpha /2$. Thus, solitons with internal frequency
[see Eq. (\ref{cos})] $\omega _{\mathrm{sol}}=\cos \theta <\cos \left(
\alpha /2\right) $ can be pinned to the interface between two BGs with phase
jump $\alpha $. The limit of $\omega _{\mathrm{sol}}=\cos \left( \alpha
/2\right) $ exactly corresponds to the linear mode pinned to the same
interface, as per Eq. (\ref{L=0}).

\subsubsection{The asymmetric system}

Composite flat-top solitons can be constructed for the asymmetric system
too, following the pattern of ansatz (\ref{soliton}), but it is then
necessary to introduce two different offsets $\xi $ in the left and right
semi-infinite gratings, which makes the composite analytical solution very
cumbersome. However, the interaction of the BG soliton with the weak
asymmetric interface, which is characterized by small $\alpha \ll \pi /2$
and $L\ll \sin \theta $ in Eq. (\ref{kappa2}), can be analyzed by means of
the simplest (adiabatic) version of the perturbation theory, which uses the
Hamiltonian term (\ref{int}), accounting for the Bragg reflectivity, to
generate an effective potential of the interaction of a BG soliton with the
interface, $W_{\mathrm{int}}$, while the perturbation of the soliton's shape
is disregarded \cite{KM}. A straightforward analysis yields the following
asymptotic expression for the interaction force acting on the soliton with
the center located at point $\Xi $ far to the left from the interface, $-\Xi
\sin \theta \gg 1$:%
\begin{equation}
F_{\mathrm{int}}=-\frac{\partial W_{\mathrm{int}}}{\partial \Xi }\approx
\frac{8}{3}\left( \sin ^{2}\theta \right) \left( 4L\sin \theta -\alpha
^{2}\right) \exp \left( -2\left( \sin \theta \right) \Xi \right) .  \label{F}
\end{equation}%
The soliton is captured by the weak interface if the the remote soliton is
attracted to it, i.e., $F_{\mathrm{int}}>0$. Thus, it follows from Eq. (\ref%
{F}) that the capture does not occur if the soliton's amplitude is not large
enough, $\sin \theta <\alpha ^{2}/\left( 4L\right) $. Because $\sin \theta $
cannot exceed $1$, the final conclusion is that the interface \emph{cannot}
trap any BG soliton under condition%
\begin{equation}
L<L_{\min }^{\mathrm{(sol)}}=\alpha ^{2}/4.  \label{Lmin2}
\end{equation}%
This prediction is qualitatively similar to that for the existence condition
of the linear DM, given by Eq. (\ref{Lmin}) with $n=0$, the difference being
that, for $\alpha \rightarrow 0$, $L_{\min }^{\mathrm{(sol)}}\sim \alpha
^{2} $ is much smaller than $L_{\min }^{\mathrm{(DM)}}\sim \alpha $.

\section{Numerical results for gap solitons}

Solitons in the symmetric system (with the phase shift $\alpha $)\ were
obtained above in the implicit analytical form, based on Eqs. (\ref{soliton}%
)-(\ref{eigen2}), Here, we focus on constructing a family of fundamental
solitons [whose linear limit correspond to $n=0$ in Eq. (\ref{eigen-asymm})]
in the asymmetric system in a numerical form, by means of the Newton's
method. The results also include the symmetric system with $\alpha =0$ (the
one with no phase shift). To this end, the exact gap soliton, given by Eq. (%
\ref{sol}) for the uniform BG,\ was used as the initial guess. An essential
conclusion of the numerical analysis, which complies with the similar
property of the symmetric system [represented by the above-mentioned
correspondence between Eqs. (\ref{eigen2}) and (\ref{eigen}), as well as
between (\ref{ln}) and (\ref{L=0})], is that nonlinear DMs, obtained as a
continuation of their linear counterparts given by Eqs. (\ref{U(x)})-(\ref%
{eigen-asymm}), are tantamount to gap solitons pinned to the gapless layer.
Stability of the so found solitons was then tested by means of direct
simulations of their perturbed evolution, which were run up to $t=10000$,
using the split-step fast-Fourier-transform algorithm, with absorbers set at
edges of the integration domain.

\begin{figure}[tbp]
\centering
\subfigure[]{\includegraphics[width=3in]{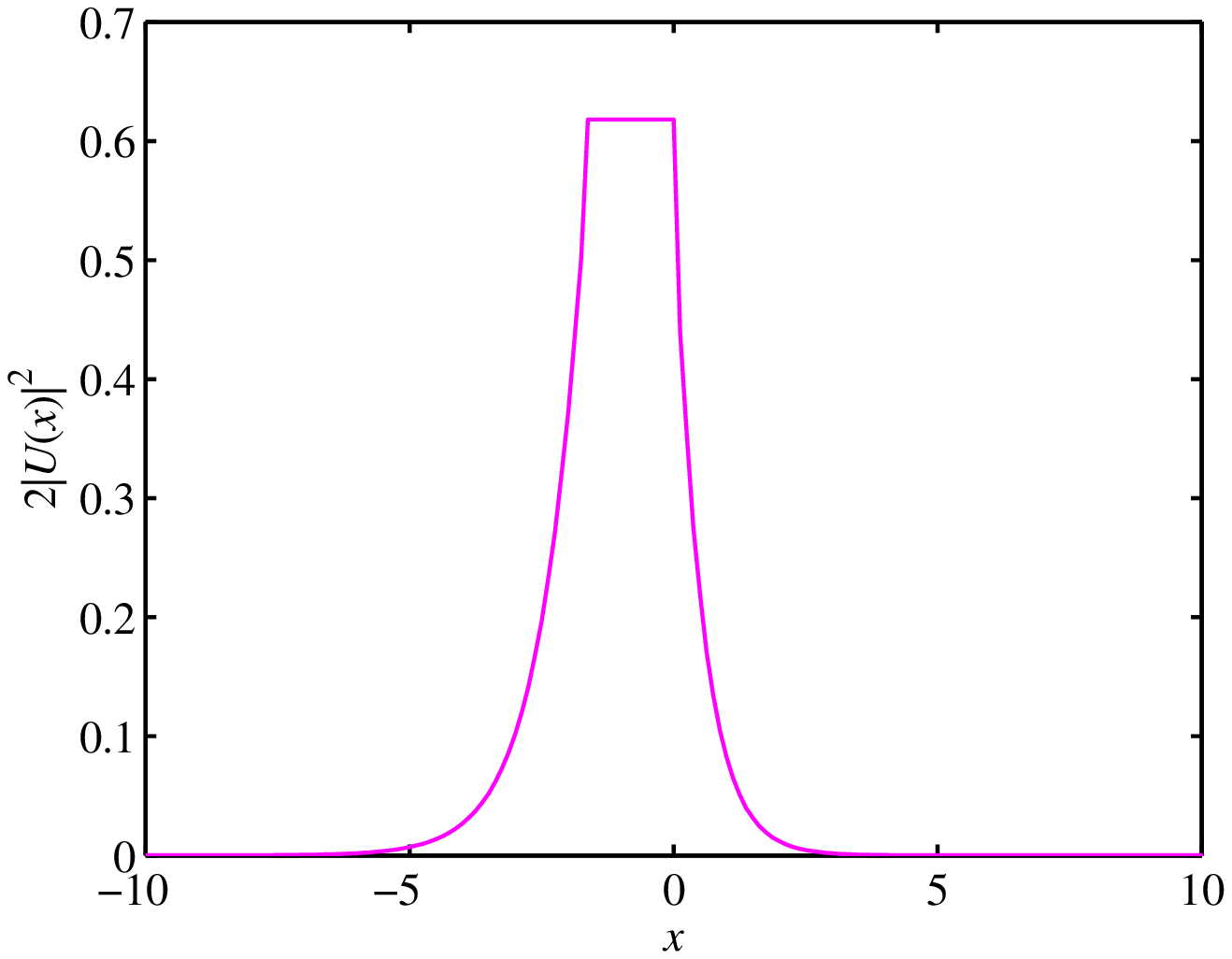}}\subfigure[]{%
\includegraphics[width=3in]{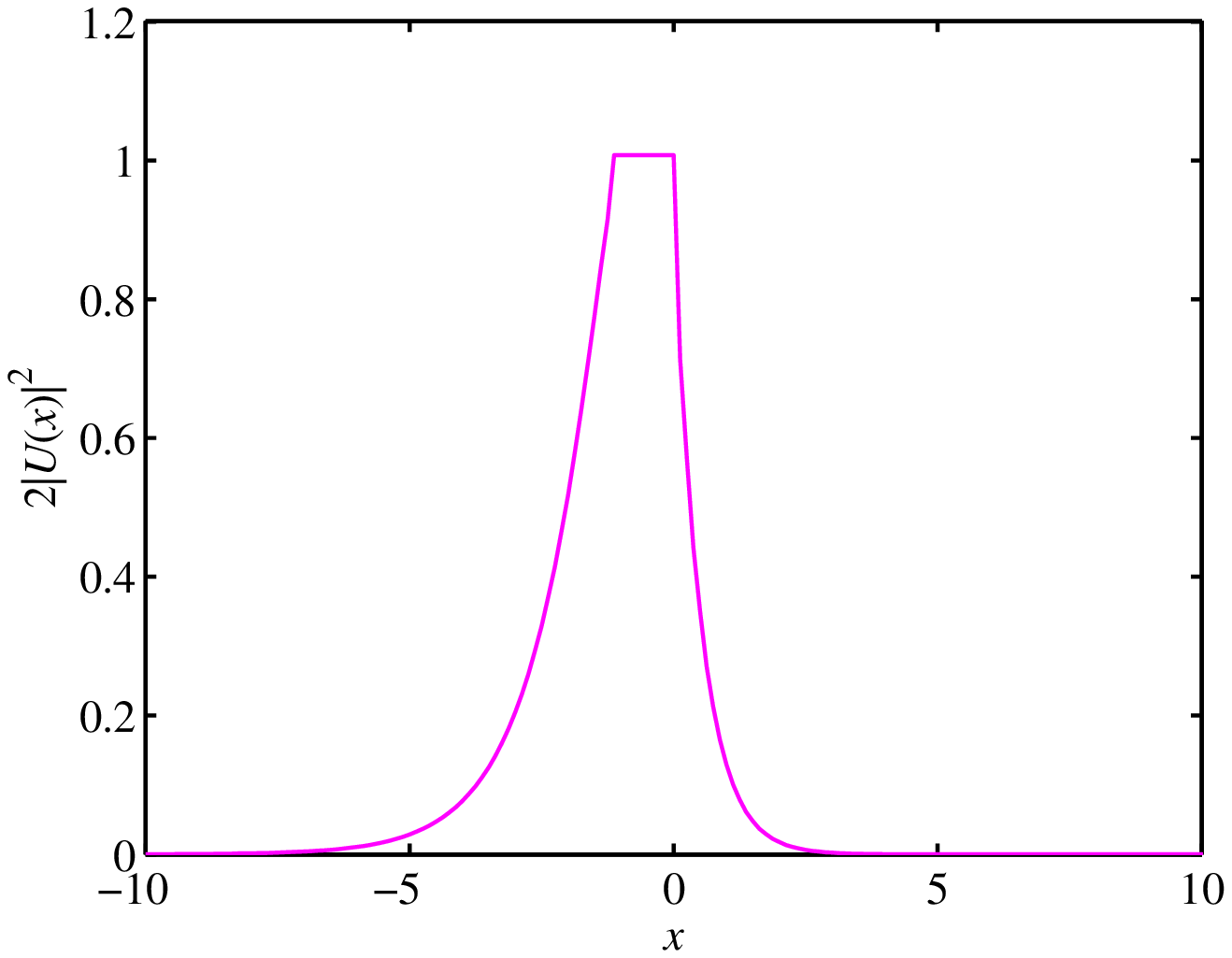}}
\caption{(Color online) Example of intensity profiles of numerically found
solitons in the asymmetric system. (a) A stable soliton for $\protect\alpha =%
\protect\pi /4$, $L=1.625$, $\protect\omega =0.2$. (b) An unstable soliton
for $\protect\alpha =\protect\pi /3$, $L=1.125$, $\protect\omega =-0.2$}
\label{examples}
\end{figure}

Typical examples of numerically found stable and unstable composite solitons
in the asymmetric system are presented in Fig. \ref{examples}. The asymmetry
is clearly featured by shapes of the solitons, cf. the symmetric ones in
Figs. \ref{figure6}.

\begin{figure}[tbp]
\centering\subfigure[]{\includegraphics[width=3in]{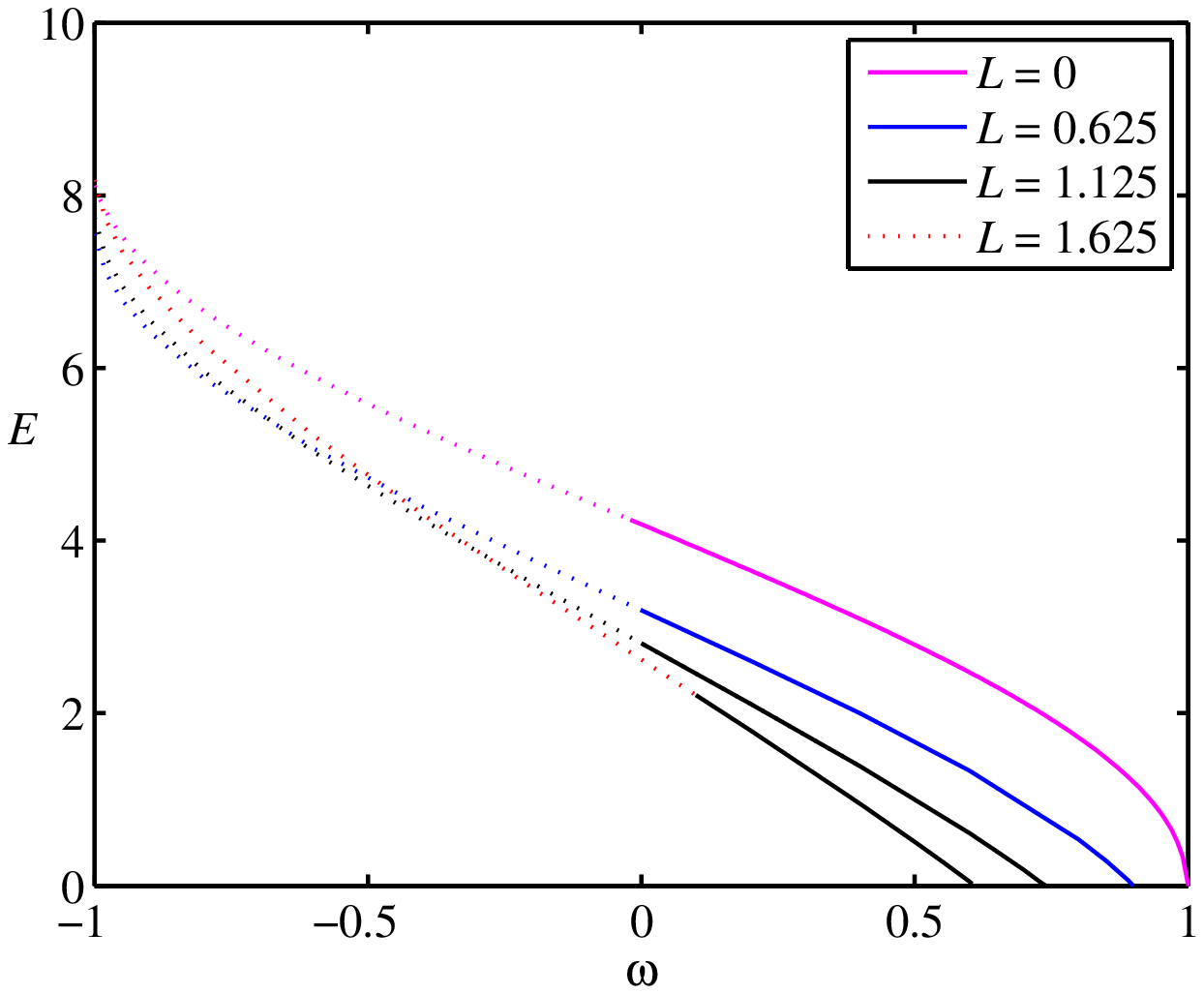}}%
\subfigure[]{\includegraphics[width=3in]{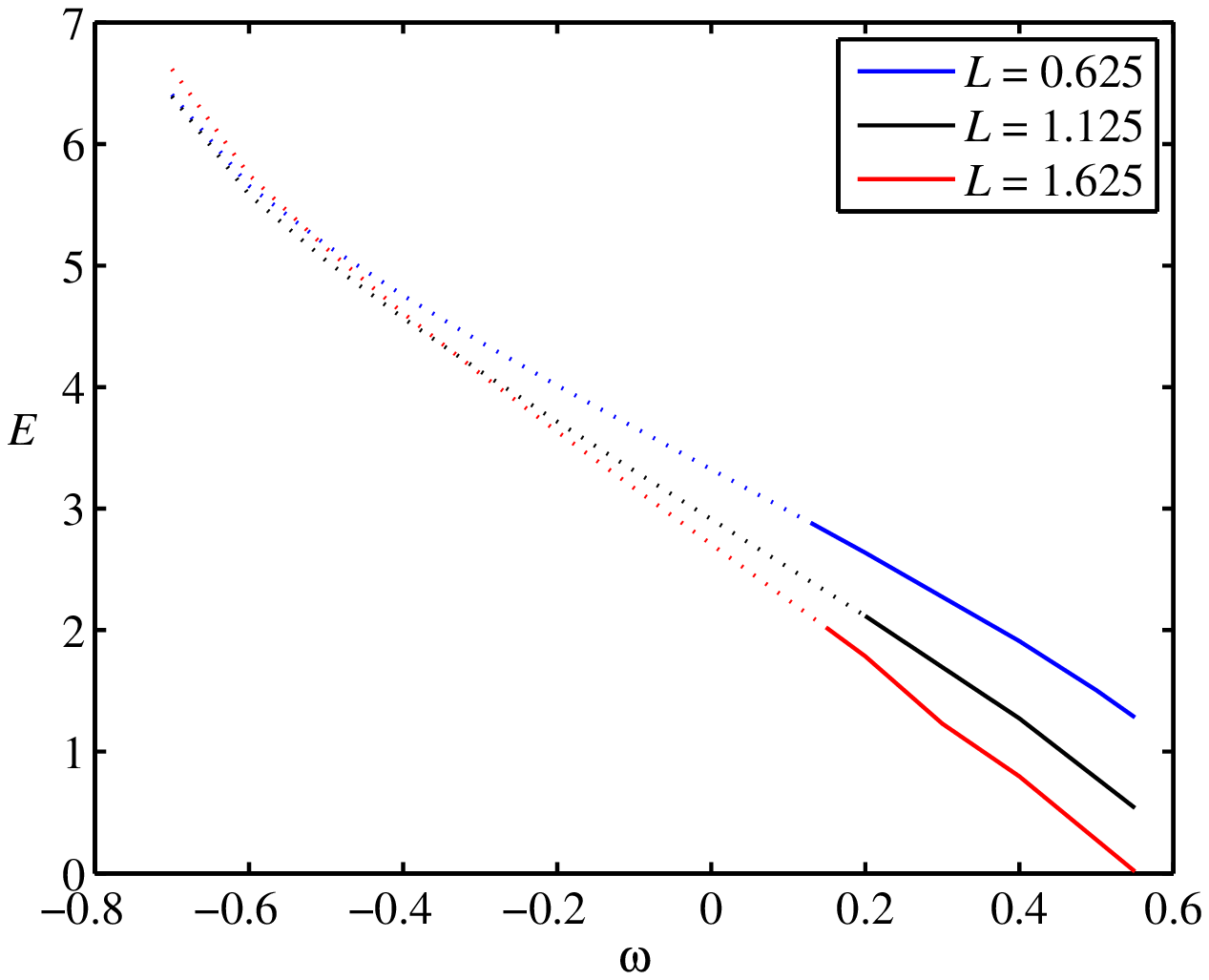}} \subfigure[]{%
\includegraphics[width=3in]{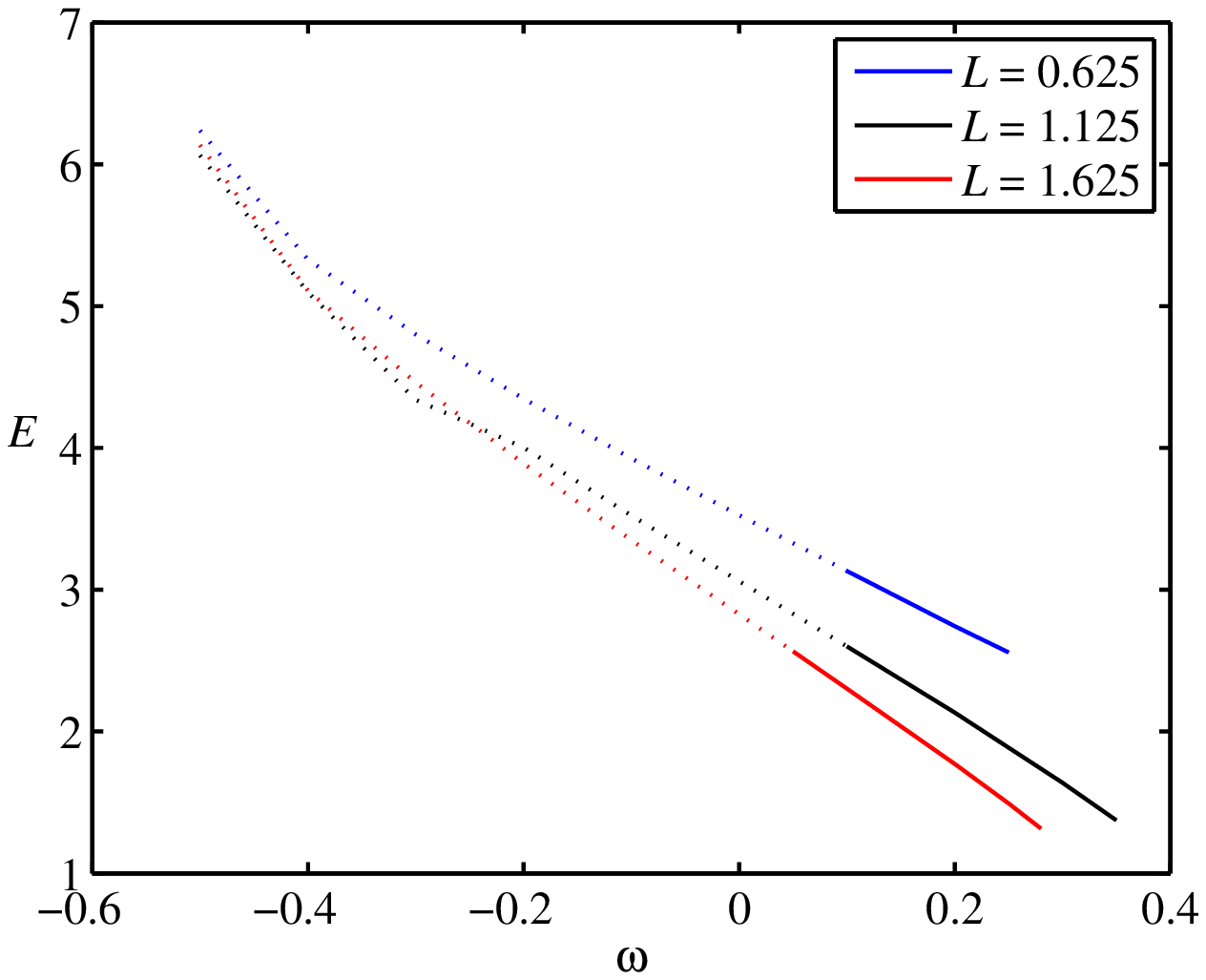}}
\caption{(Color online) Energy $E$ of the gap solitons versus $\protect%
\omega $ in the asymmetric system for different $L$. (a) $\protect\alpha =0$%
; (b) $\protect\alpha =\protect\pi /4$; (c) $\protect\alpha =\protect\pi /3.$
Continuous and dotted segments denote stable and unstable solutions,
respectively. In (a), the curve for $L=0$ represents the exact solution (%
\protect\ref{sol}), with $E=(8/3)\arccos \protect\omega $, as per Eqs. (%
\protect\ref{cos}) and (\protect\ref{Esol}), the boundary between the stable
and unstable segments corresponding to Eq. (\protect\ref{crit}). In panels
(b) and (c), the numerical solution cannot be extended beyond bottom points
at which the branches terminate.}
\label{figure4}
\end{figure}

Soliton families are characterized by the dependence of the total energy, $E$
[see Eq. (\ref{E})], on frequency $\omega $, which is shown in Fig. \ref%
{figure4} for different fixed values of $L$. This figure also designates
stable and unstable portions of the solution branches. Note that values of $%
\omega $ at $E\rightarrow 0$ in the symmetric system with $\alpha =0$ [panel %
\ref{figure4}(a)], which, obviously, correspond the linear DM, exactly agree
with the respective eigenvalues (\ref{4-L}). On the other hand, the fact
that the solution branches cannot be extended up to $E=0$, i.e., to the
linear limit, at $\alpha >0$, and become progressively shorter with the
increase of $\alpha $ and decrease of $L$, as seen in panels \ref{figure4}%
(b,c), is explained by the nonexistence of the linear DM in the asymmetric
system for $L$ too small, according to Eq. (\ref{Lmin}), as well as by the
nonexistence of the full nonlinear solutions at small $L$, according to Eq. (%
\ref{Lmin2}). In terms of the numerical solution, the limitation on the
existence of the localized modes imposed by the asymmetry is more severe
than predicted by Eq. (\ref{Lmin}), because exactly at the existence limit
given by this equation, $L=\alpha /\left( 2\cos \alpha \right) $, the mode's
intrinsic frequency coincides with the edge of the bandgap, i.e., the mode
is still delocalized and cannot be found in the numerical form in a finite
domain.

\begin{figure}[tbp]
\centering\subfigure[]{\includegraphics[width=3in]{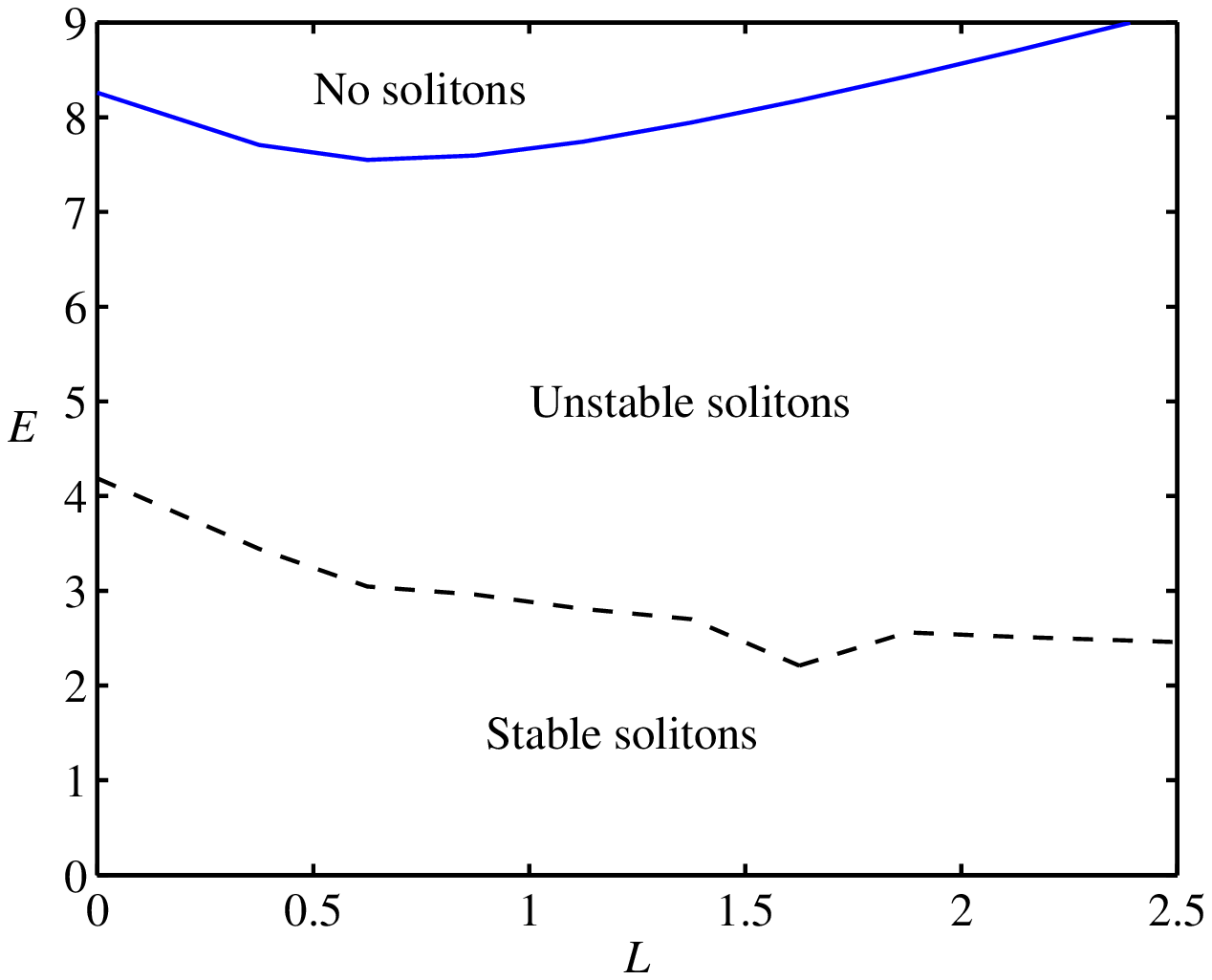}}%
\subfigure[]{\includegraphics[width=3in]{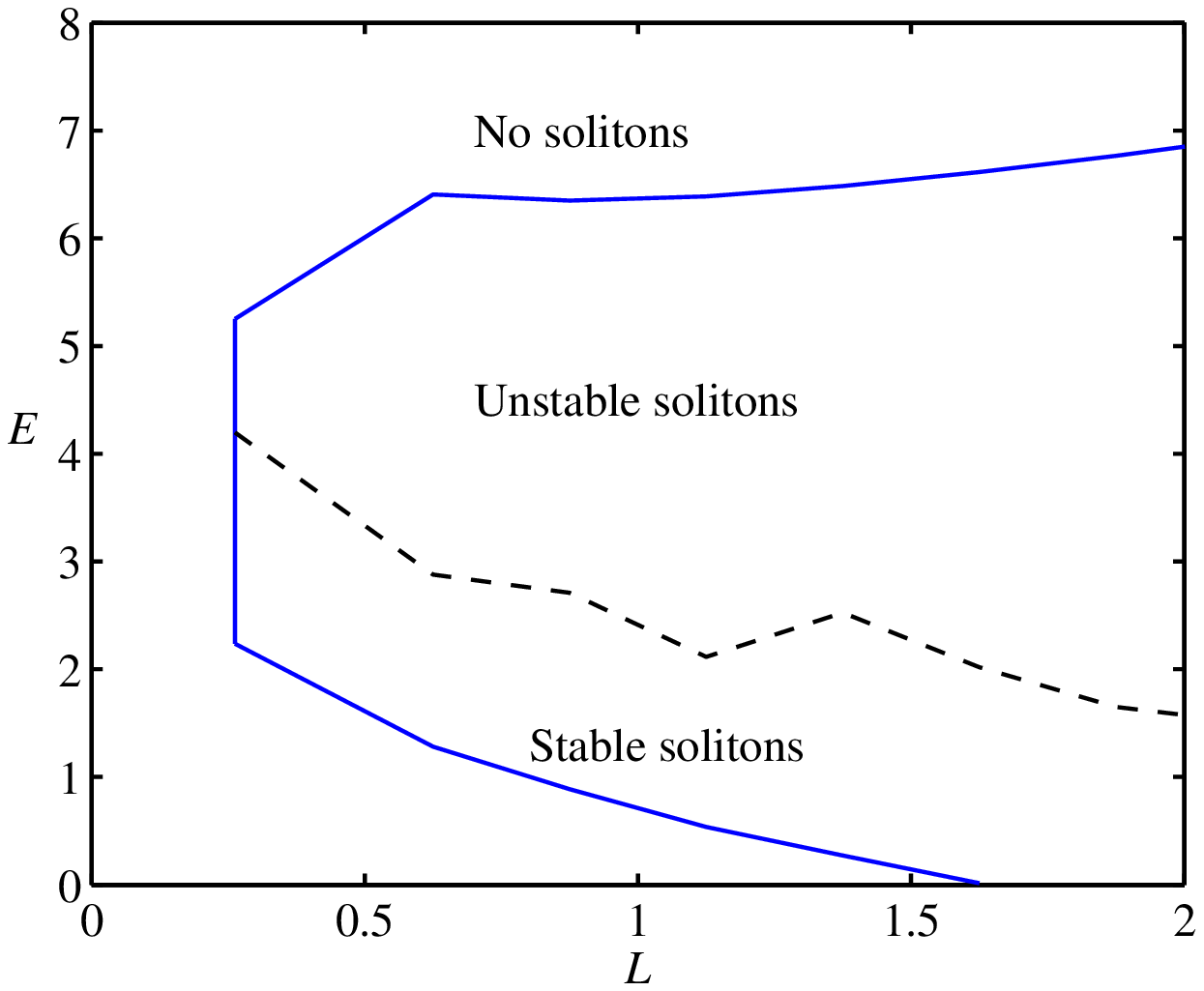}} \subfigure[]{%
\includegraphics[width=3in]{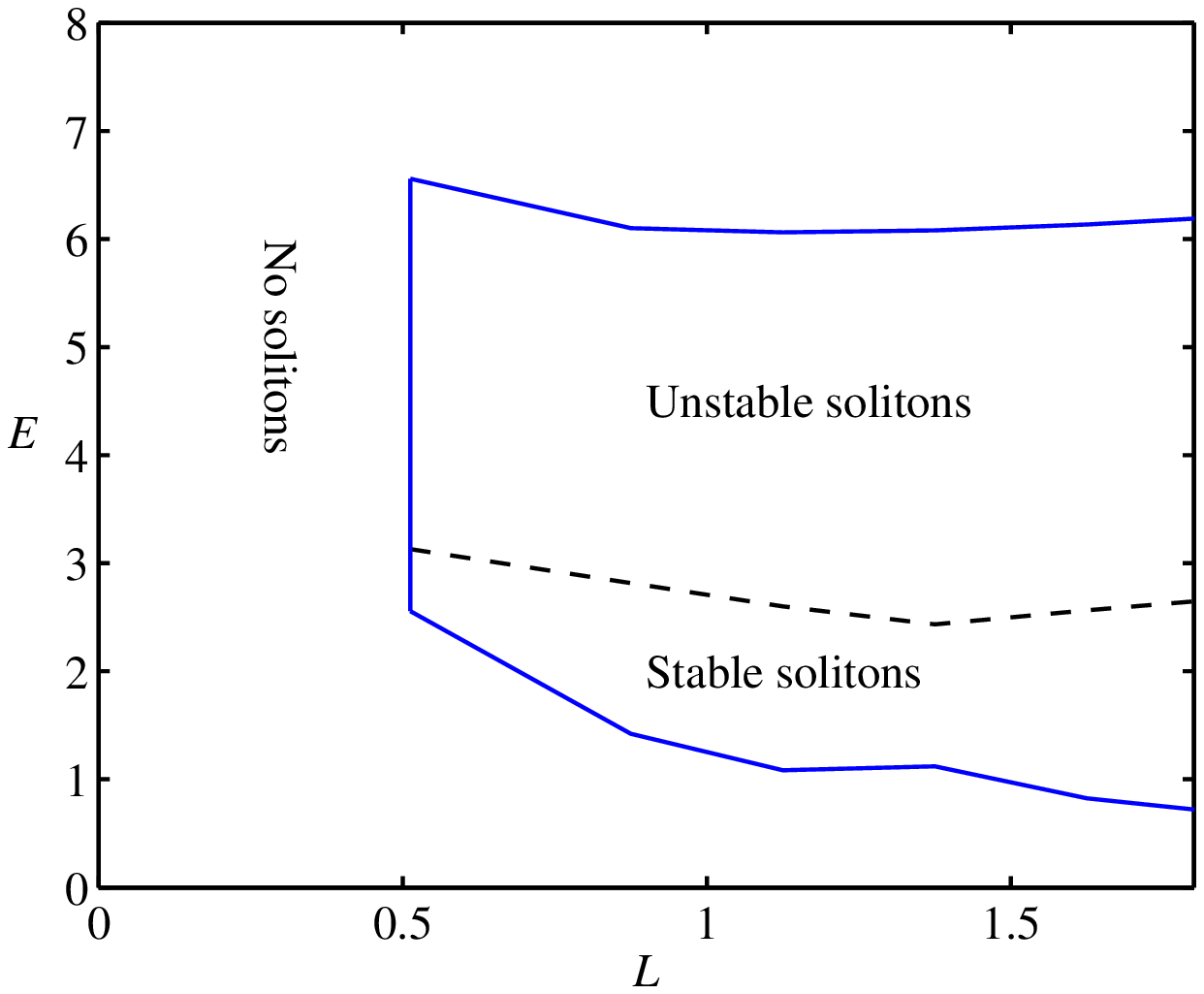}}
\caption{(Color online) Existence and stability regions for the pinned gap
solitons in the plane of $\left( L,E\right) $ of the asymmetric system at
different fixed values of $\protect\alpha $: (a) $\protect\alpha =0$ (the
symmetric system); (b) $\protect\alpha =\protect\pi /4$; (c) $\protect\alpha %
=\protect\pi /3.$ In the case of $\protect\alpha =0$, the results for $L=0$
correspond to Eqs. (\protect\ref{Esol}) and (\protect\ref{crit}).}
\label{figure7}
\end{figure}

Figure \ref{figure7} summarizes the numerical results by depicting stability
boundaries for the solitons in the plane of $\left( L,E\right) $ for
different fixed values of $\alpha $ in Eq. (\ref{kappa2}). It was found that
the existence and stability areas vanish at $\alpha >\alpha _{\mathrm{cr}%
}\approx \pi /2.8$, which is explained by the above-mentioned trend to the
disappearance of the localization modes with the increase of asymmetry $%
\alpha $ [obviously, no localized modes may exist at $\alpha =\pi /2$, when
bandgap \ (\ref{bandgap}) with $\kappa =\cos \alpha $ shrinks to zero].
Further, in accordance with the above analytical predictions [see Eqs. (\ref%
{Lmin}) and (\ref{Lmin2})], gap solitons cannot be pinned by the asymmetric
layer whose thickness, $L$, is too small, while there is no such limitation
for the symmetric layer, with $\alpha =0$. Finally, following the pattern of
the gap solitons in the uniform BG [see Eqs. (\ref{crit}) and (\ref{Esol})],
the pinned solitons tend to destabilize with the increase of the energy, and
they do not exist at very large values of the energy.

\begin{figure}[tbp]
\centering
\includegraphics[width=3in]{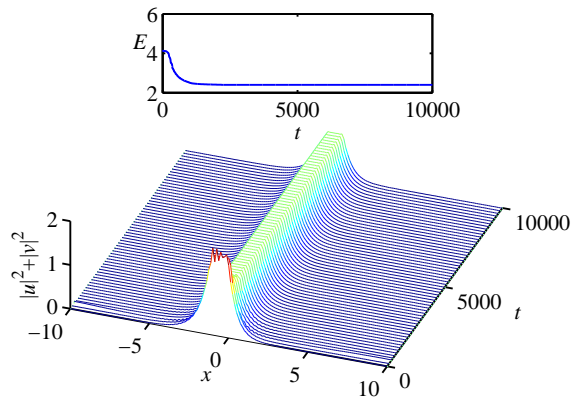}
\caption{(Color online) A typical example of the evolution of an unstable
soliton in the asymmetric system, at $\protect\omega =-0.3$, $L=1.125$, $%
\protect\alpha =\protect\pi /4$. The top panel shows the drop of the
soliton's total energy, due to radiation loss in the course of the
transition to the stable shape. The eventually established soliton can be
identified as one with $\protect\omega =0.2$, which lies at the edge of the
stability interval in Fig. \protect\ref{figure4}(c). }
\label{figure8}
\end{figure}

Evolution of unstable solitons was investigated by means of direct
simulations. As shown in Fig. \ref{figure8}, unstable solitons are not
destroyed. Instead, they shed off a part of their energy (which is
eventually eliminated by the edge absorbers) and thus transform into stable
solitons with smaller energy.

\section{Collisions of moving gap solitons with the gapless layer}

Collisions of moving BG solitons with the grating-free layer is another
natural problem in the present setting. It was addressed by means of direct
simulations, starting with a soliton initially set far from the layer, i.e.,
with the center placed at some large $x_{0}>0$, and with initial velocity $%
c<0,$\ so that it travels to the left, hitting the central layer. If the
layer attracts the soliton, the collision may result in its splitting into
reflected and transmitted pulses, plus a trapped mode. In the uniform
grating with $\kappa =1$, the initial conditions for the moving gap soliton
are taken as the exact solution (written here at the initial moment, $t=0$)
\cite{gapsol1,gapsol2}:

\begin{eqnarray}
u_{\mathrm{sol}} &=&\sqrt{\frac{2\left( 1+c\right) }{3-c^{2}}}\left(
1-c^{2}\right) ^{1/4}W(X)\mathrm{exp}\,\left[ i\phi (X)-iT\cos \theta \right]
,  \notag \\
v_{\mathrm{sol}} &=&-\sqrt{\frac{2\left( 1-c\right) }{3-c^{2}}}\left(
1-c^{2}\right) ^{1/4}W^{\ast }(X)\mathrm{exp}\,\left[ i\phi (X)-iT\cos
\theta \right] ,  \label{movsol}
\end{eqnarray}%
{\LARGE \noindent\ }%
\begin{eqnarray}
X &=&\left( 1-c^{2}\right) ^{-1/2}x,\,T=-c\left( 1-c^{2}\right) ^{-1/2}t,
\notag \\
\phi (X) &=&\frac{4c}{3-c^{2}}\mathrm{\tan }^{-1}\left\{ \tanh \left[ (\sin
\,\theta )X\right] \tan \left( \theta /2\right) \right\} ,  \label{params} \\
W(X) &=&\left( \sin \,\theta \right) \,\mathrm{sech}\left[ (\sin \,\theta
)X-i\left( \theta /2\right) \right] \,,  \notag
\end{eqnarray}%
the above solution (\ref{sol}) corresponding to $c=0$. In this section we
report results for the soliton with $\theta =0.29\pi $, which is stable in
the uniform BG.

We first consider the layer with $L=1.0$ and $\alpha =0$ (the symmetric
system). Results of the collisions with this layer are shown in Fig. \ref%
{figure9}. A slow soliton, originally moving with velocity $c=-0.1$, splits
into two parts, a reflected soliton and a trapped mode. The latter one keeps
$23\%$ of the total energy of the incident soliton. The incoming soliton
with $c=-0.5$ splits into three parts: a reflected one, a trapped mode, and
a weak transmitted radiation jet. In this case, the trapped mode keeps $13\%$
of the total energy. Lastly, a fast incident soliton with $c=-0.9$ splits
into two parts, a transmitted one, and a weak reflected radiation jet.
\begin{figure}[tbp]
\centering\subfigure[]{\includegraphics[width=3in]{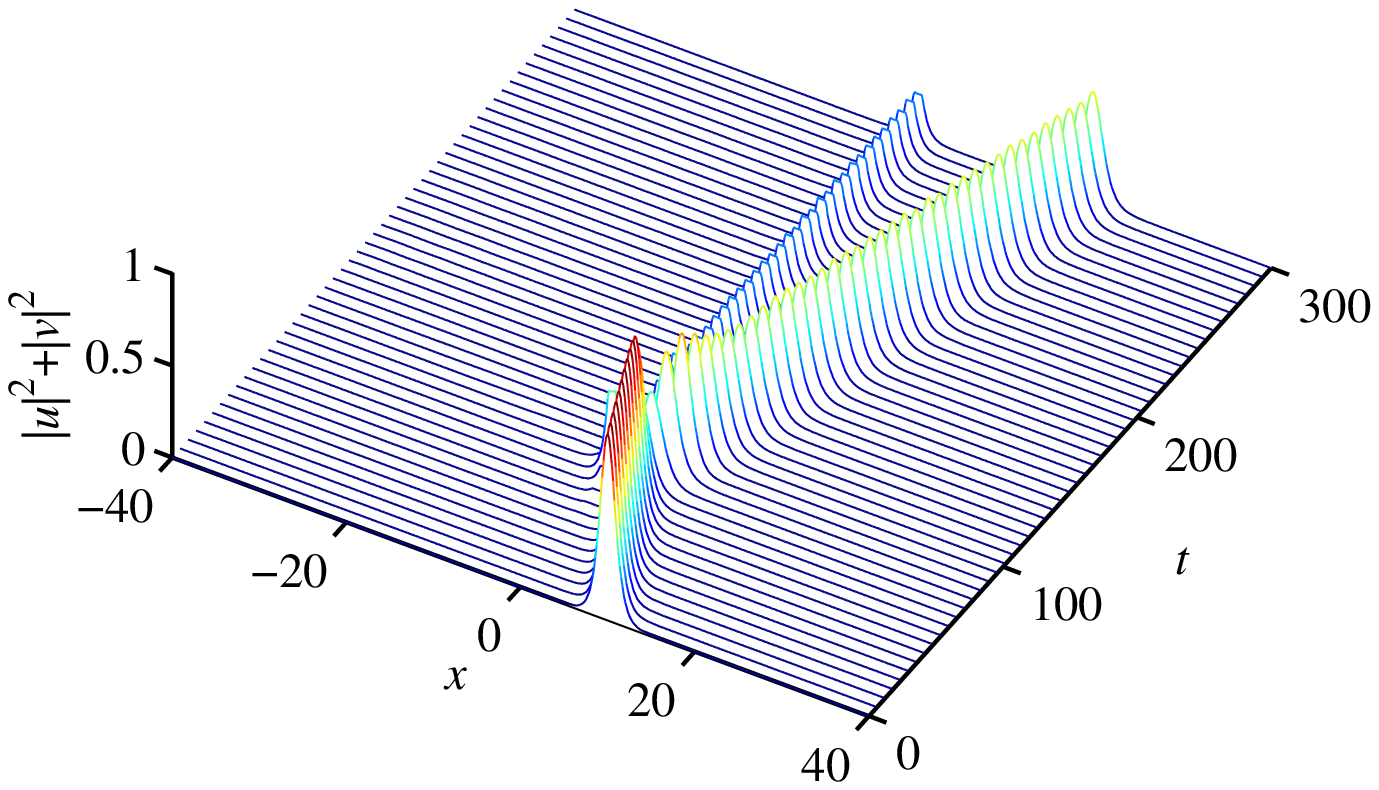}}%
\subfigure[]{\includegraphics[width=3in]{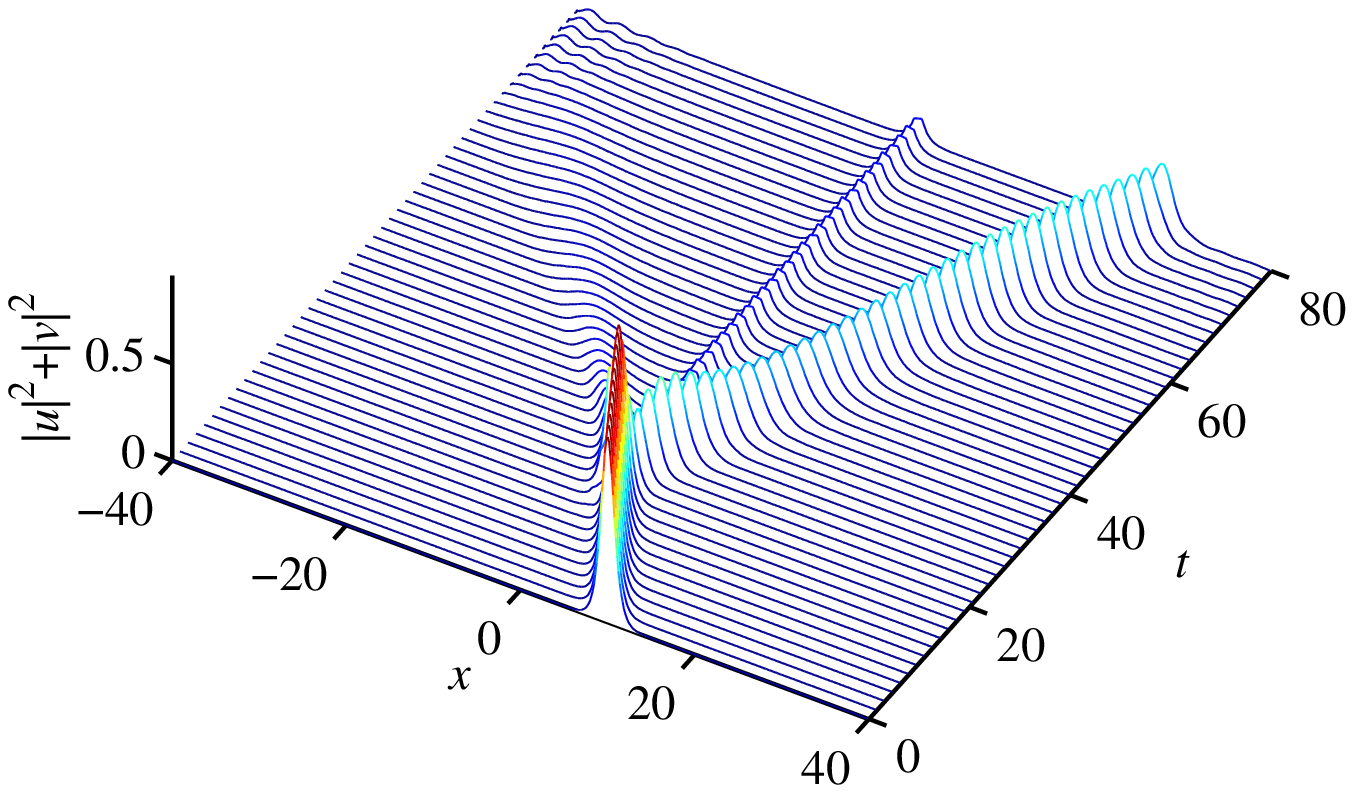}} \subfigure[]{%
\includegraphics[width=3in]{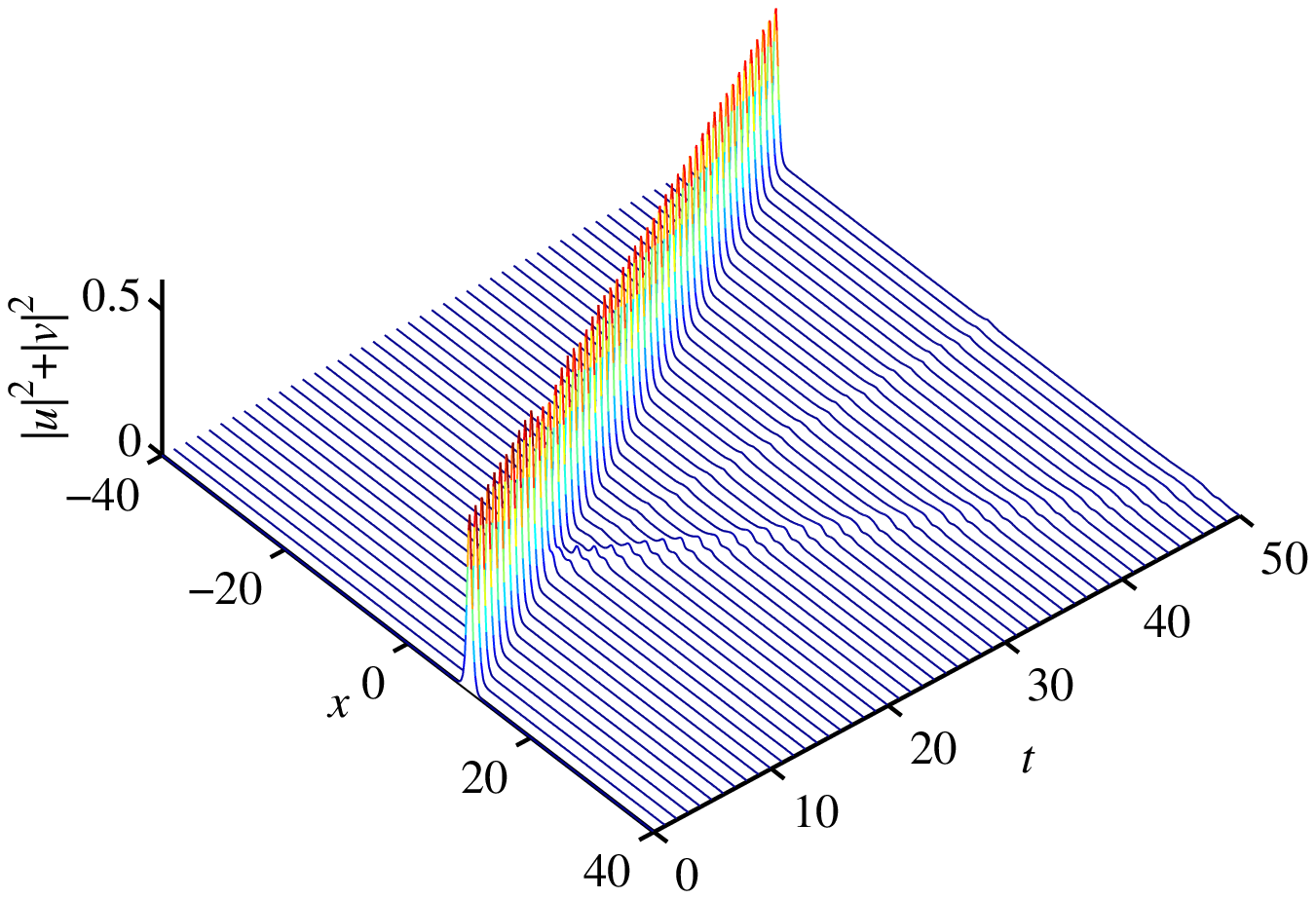}}
\caption{(Color online) Collisions of moving gap solitons with the gapless
layer in the symmetric system, with $\protect\alpha =0$. (a) A slow gap
soliton with velocity $c=-0.1$ splits into a reflected pulse and a trapped
mode. (b) A gap soliton with $c=-0.5$ splits into a reflected pulse, a
trapped mode, and radiation. (c) A fast gap soliton with $c=-0.9$ splits
into a transmitted soliton and a radiation jet.}
\label{figure9}
\end{figure}

\begin{figure}[tbp]
\centering\subfigure[]{\includegraphics[width=3in]{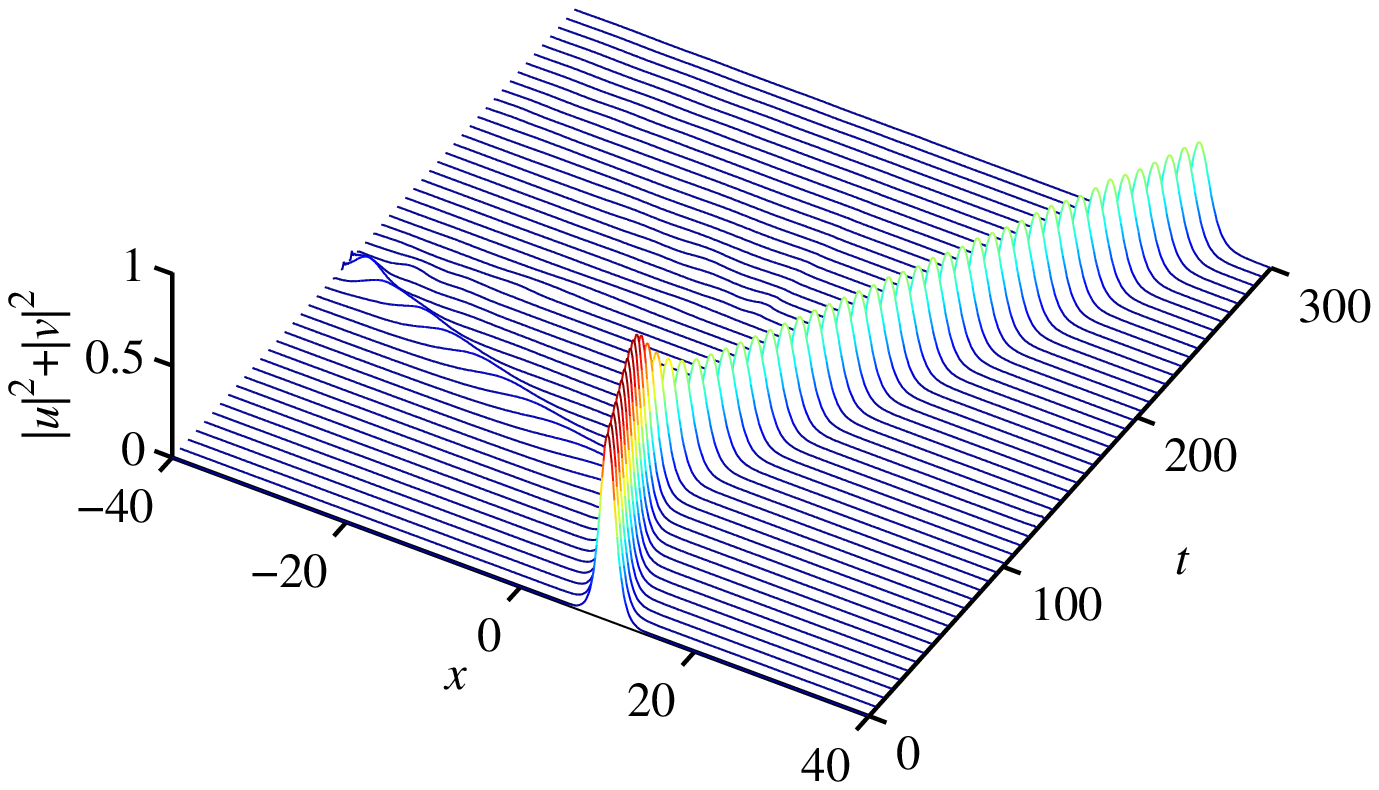}}%
\subfigure[]{\includegraphics[width=3in]{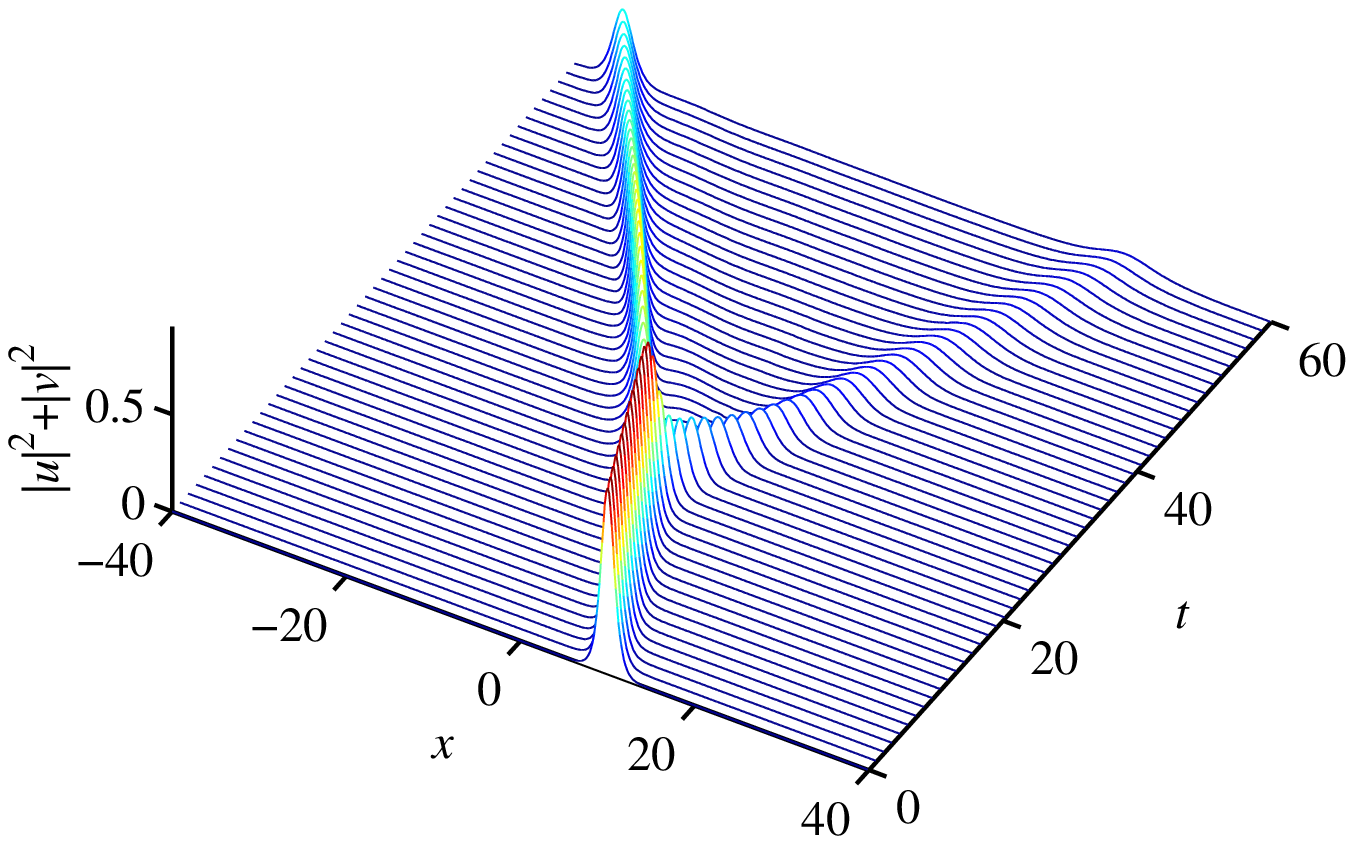}} \subfigure[]{%
\includegraphics[width=3in]{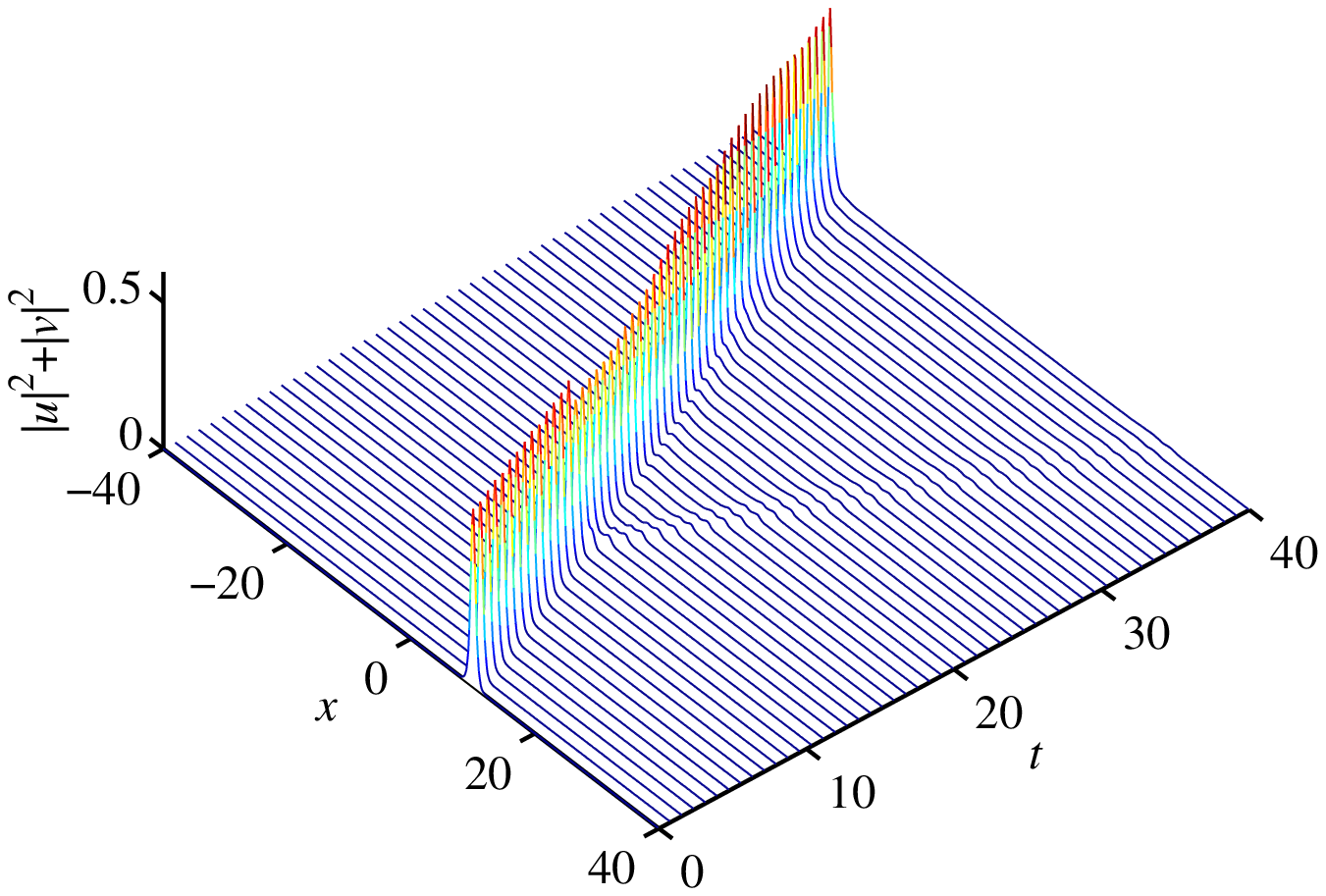}}
\caption{(Color online) Collisions in the asymmetric system, with $L=1$ and $%
\protect\alpha =\protect\pi /3$. (a) A slow soliton with $c=-0.1$ is
reflected. (b) A soliton with $c=-0.5$ splits into a transmitted one and a
reflected radiation jet. (c) A fast soliton with $c=-0.9$ is transmitted.}
\label{figure10}
\end{figure}

As shown in Fig. \ref{figure10}, the difference in the case of the collision
of the moving soliton with the gapless layer in the asymmetric system, with
sufficiently large values of $\alpha $, is that the trapped mode does not
emerge as a result. This observation is explained by the reduction and
disappearance of the stability region for the pinned solitons with the
growth of $\alpha ,$ as seen in Fig. \ref{figure7}. In particular, Fig. \ref%
{figure10} demonstrates that, at $\alpha =\pi /3$, the slow incident soliton
with $c=-0.1$ is reflected, and faster ones, with $c=-0.5$ and $c=-0.9$, are
transmitted. In addition to these outcomes, radiation jets are generated
too. Note that the transmission is facilitated, in this case, by the fact
the incident soliton is moving from the domain with the stronger Bragg
reflectivity to the one with the weaker reflectivity, according to Eq. (\ref%
{kappa2}).

\section{Conclusion}

We have introduced two models of the Kerr-nonlinear optical medium, which
include two semi-infinite BGs (Bragg gratings), separated by the
grating-free (gapless) layer of width $L$. In the symmetric model, the
bandgap in both gratings is the same, while a phase shift between them, $%
\alpha $, is possible. The other model is asymmetric, with different widths
of the bandgaps on both sides of the gapless layer, due to different Bragg
reflectivities . The model can be implemented in the temporal and spatial
domains alike. The attraction of the fields to the gapless layer gives rise
to a DM\ (defect mode) in the linear version of the systems, and pinned
composite gap solitons in the nonlinear one. However, in the asymmetric
system the attraction competes with repulsion from the reflectivity jump,
therefore the DM and pinned solitons do not exist if $L$ is too small in
this system. Both the linear and nonlinear modes are constructed
analytically and numerically in both systems, in exact or approximate forms.
The stability region for the pinned solitons in the asymmetric system is
identified by means of systematic simulations, demonstrating the shrinkage
and disappearance with the increase of $\alpha $, and disappearance at small
$L$ too. Collisions of freely moving solitons with the gapless layer were
also studied, suggesting a possibility of the creation of the trapped mode,
i.e., a pulse of standing light, as a result of the collision.

The work can be naturally extended in other directions. In particular, while
the present analysis was focused on the fundamental trapped modes, it may be
interesting to consider higher-order ones, corresponding to $n\geq 1$ in
Eqs. (\ref{eigen}), (\ref{Ln}), (\ref{eigen-asymm}), (\ref{Lmin}), and (\ref%
{eigen2}), which exist if $L$ is large enough. Another relevant extension is
to address a similar model with other types of nonlinear terms, such as
quadratic or resonant (two-level) ones.

\section{Acknowledgment}

This work was supported by the Thailand Research Fund through grant No.
RSA5780061.

\end{document}